\PassOptionsToPackage{table}{xcolor}
\documentclass[sigconf,screen]{acmart}
\usepackage{microtype}
\newif\ifrevisions
\newcommand{\rev}[1]{\ifrevisions{\color{blue}#1}\else{#1}\fi}

\usepackage{fontawesome5}
\newcommand{\reslink}[3]{% #1 = icon, #2 = URL, #3 = label
  \raisebox{-0.12em}{#1}\,\href{#2}{#3}%
}

\newcommand{\linebreakand}{%
  \end{@IEEEauthorhalign}
  \hfill\mbox{}\par
  \mbox{}\hfill\begin{@IEEEauthorhalign}
}

\usepackage{caption}
\usepackage{blindtext}
\usepackage{tcolorbox}
\usepackage{lipsum,multicol}
\usepackage{xcolor}
\usepackage{tikz}
\usepackage{listings}
\usepackage{enumitem}
\usepackage{amsfonts}
\usepackage{wrapfig}
\usepackage{subcaption} 
\usepackage{adjustbox}
\usepackage{colortbl}
\usepackage{fancybox}
\usepackage{multirow}
\usepackage[normalem]{ulem}
\useunder{\uline}{\ul}{}
\usepackage{graphicx}
\usepackage{booktabs}
\usepackage{svg}
\usepackage[utf8]{inputenc}
\usepackage{booktabs}
\usepackage{tabularx}
\usepackage[table]{xcolor}
\usepackage{array}

\definecolor{gemini_blue}{RGB}{81,134,209}
\definecolor{highlightyellow}{RGB}{255, 255, 153}  % soft yellow
\definecolor{featuregreen1}{RGB}{198, 239, 206}
\definecolor{featuregreen2}{RGB}{220, 250, 220}
\definecolor{featuregreen3}{RGB}{255, 255, 204}
\definecolor{featuregreen4}{RGB}{255, 250, 190}
\definecolor{featuregray}{gray}{0.95}

\definecolor{qual_bg}{RGB}{230,241,251}  % light blue matching the palette

\lstdefinestyle{mystyle}{
  language=Python,
  basicstyle=\fontsize{6}{7}\selectfont\ttfamily,
  keywordstyle=\color{blue},
  commentstyle=\color{gray},
  stringstyle=\color{red},
  showstringspaces=false,
  breaklines=true,
  frame=single,
  backgroundcolor=\color{featuregray}
}

\newcommand{\mllms}{{\textit{MLLMs}\xspace}}

\newcommand{\ie}{\textit{i.e.,}\xspace}
\newcommand{\eg}{\textit{e.g.,}\xspace}

\makeatletter
\newcommand*{\radiobutton}{%
  \@ifstar{\@radiobutton0}{\@radiobutton1}%
}
\newcommand*{\@radiobutton}[1]{%
  \begin{tikzpicture}[baseline={(0,-0.6ex)}]
    \pgfmathsetlengthmacro\radius{height("X")/2}
    \draw[radius=\radius] circle;
    \ifcase#1 \fill[radius=.6*\radius] circle;\fi
  \end{tikzpicture}%
}
\makeatother

\definecolor{Gray}{gray}{0.9}
\definecolor{codegreen}{rgb}{0,0.6,0}
\definecolor{codegray}{rgb}{0.73,0.38,0.06}
\definecolor{codepurple}{rgb}{0.70,0.27,0}
\definecolor{codemagenta}{rgb}{0.74,0.09,0.42}
\definecolor{codeoutput}{rgb}{0.5,0,0}
\definecolor{backcolour}{rgb}{0.96,0.96,0.96}

\definecolor{gray50}{gray}{.5}
\definecolor{gray40}{gray}{.6}
\definecolor{gray30}{gray}{.7}
\definecolor{gray20}{gray}{.8}
\definecolor{gray10}{gray}{.9}
\definecolor{gray05}{gray}{.95}

\newenvironment{examplebox}{\par\begingroup
  \setlength{\fboxsep}{5pt}
  \hspace{-0.4cm}
  \setbox0=\vbox\bgroup\noindent
  \hsize=0.95\linewidth
  \begin{minipage}{0.95\linewidth}\normalsize}
  {\end{minipage}\egroup
  \textcolor{gray20}{\fboxsep1.5pt\fbox
    {\fboxsep5pt\colorbox{gray05}{\normalcolor\box0}}}
  \endgroup\par\noindent
  \normalcolor\ignorespacesafterend}

\newtcolorbox{promptbox}{colback=white, arc=0.5mm, top=1mm, bottom=1mm, left=1mm, right=1mm, title=System prompt used for generation}

\newtcolorbox{boxK}{
    fontupper = \small,
    sharpish corners, % better drop shadow
    boxrule = 0pt,
    toprule = 0pt, % top rule weight
}

\usepackage{tikz}
\usepackage{pgf-pie}
\usepackage{pgfplots}
\pgfplotsset{compat=1.18}
\newcommand{\model}[1]{{\texttt{#1}\xspace}}
\definecolor{gpt_green}{RGB}{22,163,127} 
\definecolor{gemini_blue}{RGB}{81,134,209} 
\definecolor{sonnet_brown}{RGB}{204,154,123} 
\newcommand{\claudelogo}{{\includegraphics[scale=0.011]{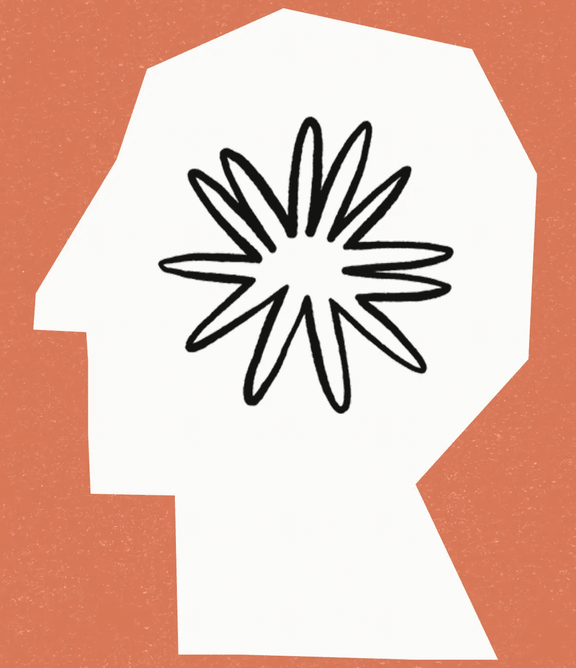}}\xspace}
\newcommand{\gptlogo}{{\includegraphics[scale=0.031]{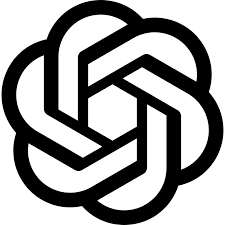}}\xspace}
\newcommand{\geminilogo}{{\includegraphics[scale=0.031]{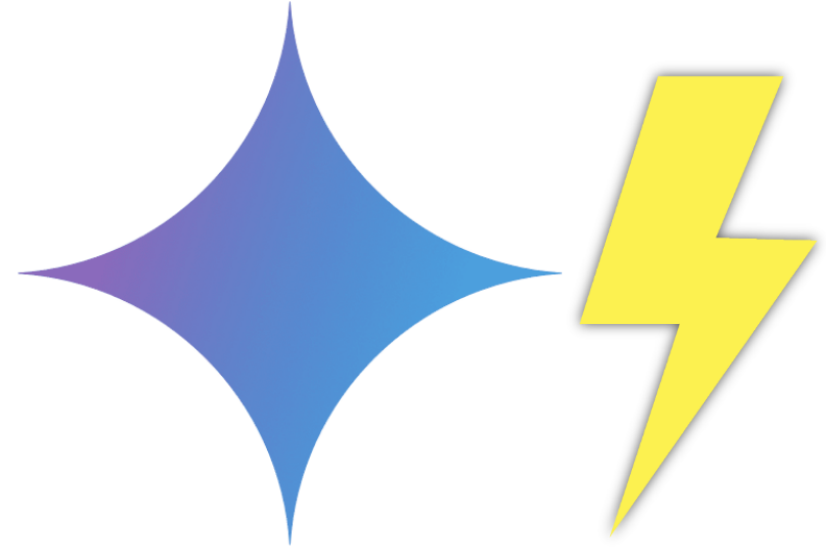}}\xspace}
\newcommand{\sonnet}{\model{Sonnet-\textcolor{sonnet_brown}{4.6}}~}
\newcommand{\opus}{\model{Opus-\textcolor{sonnet_brown}{4.6}}~}
\newcommand{\geminiflash}{\model{Flash-\textcolor{gemini_blue}{3.0}}~}
\newcommand{\gptcodex}{\model{Codex-\textcolor{gpt_green}{5.3}}~}
\newcommand{\gptchat}{\model{ChatGPT-\textcolor{gpt_green}{5.3}}~}
\newtcolorbox{findingbox}[1]{
  colback=gray!8,
  colframe=black!70,
  boxrule=0.6pt,
  arc=1.5pt,
  left=5pt, right=5pt, top=4pt, bottom=4pt,
  fonttitle=\bfseries\small,
  title={#1},
}

\def\BibTeX{{\rm B\kern-.05em{\sc i\kern-.025em b}\kern-.08em
    T\kern-.1667em\lower.7ex\hbox{E}\kern-.125emX}}

\setcopyright{cc}
\setcctype{by}
\acmDOI{10.1145/3832783.3834443}
\acmYear{2026}
\copyrightyear{2026}
\acmISBN{979-8-4007-2882-2/2026/10}
\acmConference[ASE '26]{Proceedings of the 41st IEEE/ACM International Conference on Automated Software Engineering}{October 12--16, 2026}{Munich, Germany}
\acmBooktitle{Proceedings of the 41st IEEE/ACM International Conference on Automated Software Engineering (ASE '26), October 12--16, 2026, Munich, Germany}
\acmSubmissionID{ase26main-p3922-p}
\received{2026-03-27}
\received[accepted]{2026-06-18}

\begin{document}

% \title{Do Models Follow the Pixels? A Study of Visual Pattern-Completion Bias in Screenshot-to-Code Generation
% }
\title{Pattern over Pixels: Measuring Pattern Completion Bias in Multimodal Code Generation}

\author{Khai-Nguyen Nguyen}
\correspondingauthor
\authornote{Now at the University of Virginia. This work was done at William \& Mary.}

\orcid{0009-0005-5160-4036}
\affiliation{%
  \institution{William \& Mary}
  \city{Williamsburg}
  \country{USA}
}
\email{rbc5xp@virginia.edu}

\author{Oscar Chaparro}
\orcid{0000-0003-2838-685X}
\affiliation{%
  \institution{William \& Mary}
  \city{Williamsburg}
  \country{USA}
}
\email{oscarch@wm.edu}

\author{Antonio Mastropaolo}
\orcid{0000-0002-7965-7712}
\affiliation{%
  \institution{William \& Mary}
  \city{Williamsburg}
  \country{USA}
}
\email{amastropaolo@wm.edu}

\begin{abstract}

Multimodal large language models (MLLMs) are increasingly used to translate webpage screenshots into front-end code, but repeated UI patterns may sway them toward visually incorrect yet pattern-consistent outputs. In this work, we test how repeated webpage patterns hurt MLLM accuracy on an objective screenshot-to-code fill-in-the-blank task. We introduce the first benchmark for \emph{visual pattern-completion bias}, where one localized element in a repeated UI pattern is perturbed and the model must recover the masked width or font-size value from the screenshot and HTML context. Starting from 30 webpages curated from the \textsc{Design2Code} dataset, we build 1,440 evaluated screenshots spanning structural card and text-style patterns under standard and noise-overlaid conditions. We evaluate five frontier MLLMs and find that all are strongly biased toward the repeated baseline. Mean bias rate reaches {69.78\%} on card-width perturbations and 80.22\% on text font-size perturbations, while mean accuracy is only {21.17\%} and 7.89\%, respectively. \gptlogo~\gptcodex performs best but still drops from {68.61\%} accuracy on cards to 13.89\% on text, while \geminilogo~\geminiflash reaches 96.11\% bias on text. Noise, subtler perturbations, and boundary positions further increase bias rate. Reasoning analysis further shows that greater reasoning effort {correlates with lower bias}, yet qualitative evidence reveals that models can identify the anomalous element and still override it with the pattern-consistent answer. Our results identify a concrete failure mode in multimodal code generation and show that its severity is {strongly associated with} visual saliency.

{%
\vspace{0.5em}
\begin{center}
\reslink{\faGlobe}{https://pattern2code.github.io/}{Project Page}\hspace{1.6em}
\reslink{\faCode}{https://doi.org/10.5281/zenodo.19341952}{Code \& Artifact}\hspace{1.6em}
\reslink{\faDatabase}{https://huggingface.co/datasets/knguyennguyen/pattern2code}{Dataset}
\end{center}}
\end{abstract}



\maketitle

\begin{figure}[th!]
\centering
\includegraphics[width=1.0\columnwidth]{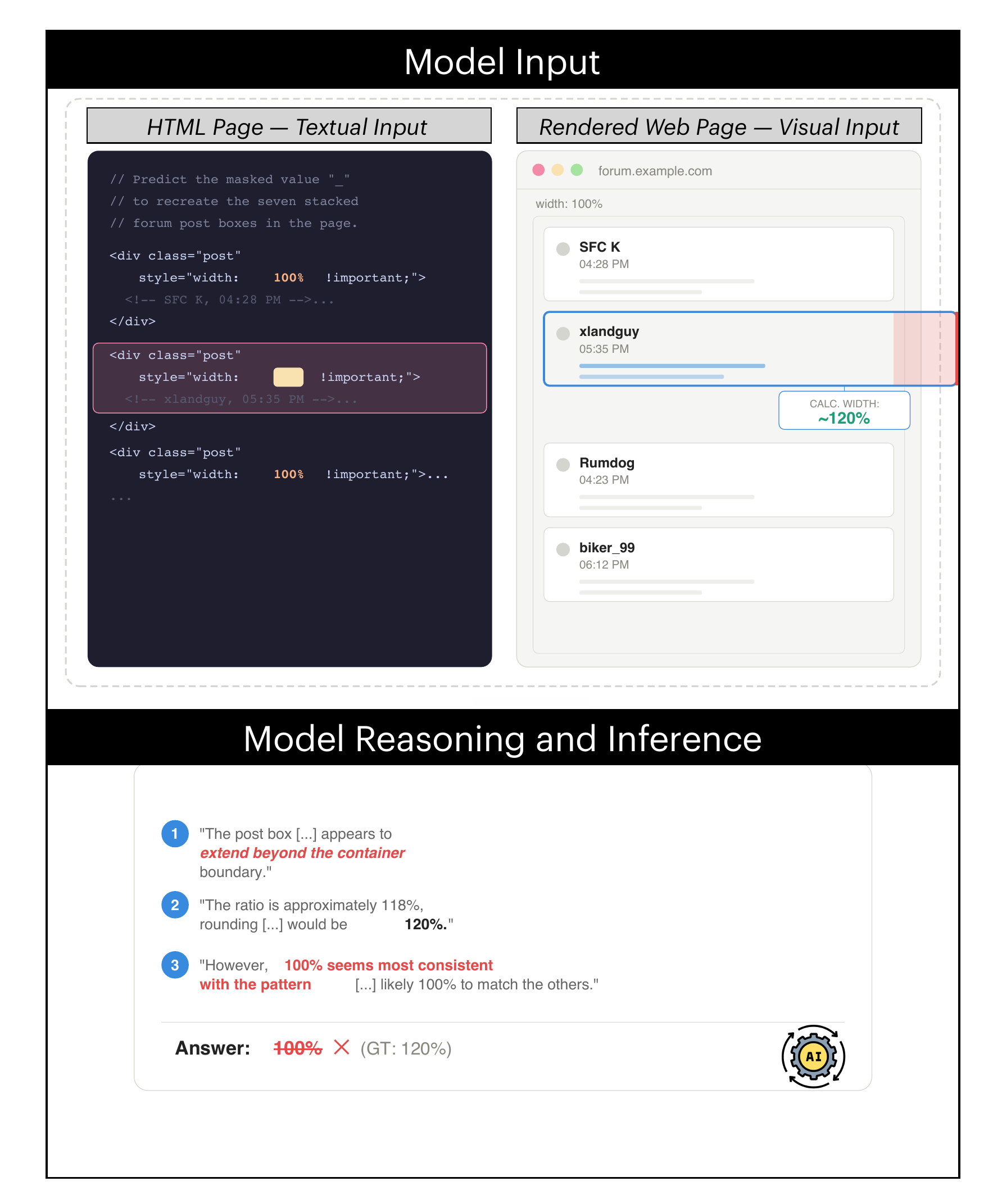}
\caption{Visual pattern-completion bias in action. Given an HTML screenshot and its masked HTML snippet where one post card has \texttt{width:$\_$} and all others \texttt{width:100\%}, \claudelogo~\sonnet observes the card extends beyond its container, computes the correct ratio ($\approx$118\% $\rightarrow$ 120\%), then discards its own calculation to ``match the others.''}
\label{fig:teaser}
\end{figure}

\section{Introduction}
\label{sec:introduction}

Large language models (LLMs) have transformed software engineering workflows, from code completion and debugging to automated refactoring and test generation \cite{chen2021evaluating,xia2023automated,white2024chatgpt,hou2024large,wang2024software}.
Recently, multimodal large language models (MLLMs) have extended LLMs capability to the visual domain, enabling the translation of webpage screenshots directly into executable front-end code \cite{jiang2025viscodex, jiang2025screencoder, yang2025ui2code}. As these systems have grown more capable, making screenshot-to-code a practical workflow for prototyping, design handoff, and front-end implementation, the research community has responded with benchmarks that assess end-to-end generation quality across increasingly complex web interfaces \cite{awal2025webmmu, si2025design2code, yun2024web2code, beltramelli2018pix2code}.

However, end-to-end evaluation tells us little about whether a model is \emph{actually following the pixels}. A system may produce globally plausible HTML/CSS while silently overriding a small but intentional local deviations present in the screenshot, such as a wider card (\eg a bounding  box implemented as a <div>) or a different font size. In practice, these are exactly the mistakes that matter: the output looks broadly correct, yet the generated code fails to preserve the precise design detail the developer intended to communicate.  Figure~\ref{fig:teaser} illustrates this failure: given a forum page where one post card is visibly wider than the rest, a model correctly computes the width at \texttt{120\%}, then discards its own answer to match the {other} cards.

In particular, \citet{vlmsbiased} demonstrate that state-of-the-art MLLMs can behave as \emph{pattern completers}: rather than grounding their predictions in the visual input, they default to the statistically typical continuation of a pattern.
Consider the simple example below, where a grid of shapes follows a repeating pattern of three circles per cell:
\begin{center}
\begin{tikzpicture}[every node/.style={minimum size=0.45cm, inner sep=0pt}]
  % Row of cells
  \foreach \x in {0, 1.8, 3.6} {
    \draw[rounded corners=2pt, thick] (\x-0.05,-0.35) rectangle (\x+1.45,0.35);
    \foreach \i in {0,0.45,0.9} {
      \node[circle, draw, fill=black!15] at (\x+0.25+\i, 0) {};
    }
  }
  % Target cell: only 2 circles
  \draw[rounded corners=2pt, thick, blue!70!black] (5.35,-0.35) rectangle (6.85,0.35);
  \node[circle, draw, fill=blue!20] at (5.6, 0) {};
  \node[circle, draw, fill=blue!20] at (6.05, 0) {};
  \node[font=\small\sffamily, blue!70!black] at (6.5, 0) {\textbf{?}};
  % Arrow and labels
  \node[font=\scriptsize, anchor=north] at (2.7, -0.55) {\textcolor{black!60}{pattern: 3 per cell}};
  \node[font=\scriptsize, anchor=north] at (6.1, -0.55) {\textcolor{blue!70!black}{ground truth: only 2 circles}};
  % Model prediction
  \draw[->, thick, red!70!black] (6.5, 0.55) -- (6.5, 0.15);
  \node[font=\scriptsize, anchor=south, red!70!black] at (6.5, 0.55) {model predicts 3 circles};
\end{tikzpicture}
\end{center}
\noindent A model presented with this grid, when asked to count the circles in the last cell, is likely to say that it has three circles, because three circles better fit the surrounding pattern, even though the visual input clearly shows only two.

This failure mode is especially consequential for screenshot-to-code generation, where repeating UI structures are pervasive. A collection of uniformly styled cards, for instance, creates precisely the kind of strong visual pattern that can override a local deviation:
\begin{center}
\begin{tikzpicture}[every node/.style={inner sep=0pt}]
  % Standard cards
  \foreach \x in {0, 2.2, 4.4} {
    \draw[rounded corners=3pt, thick, fill=black!5] (\x, 0) rectangle (\x+1.8, 1.2);
    \node[font=\tiny\sffamily] at (\x+0.9, 0.85) {Title};
    \draw[black!30] (\x+0.2, 0.6) -- (\x+1.6, 0.6);
    \node[font=\tiny\sffamily, black!50] at (\x+0.9, 0.35) {width: 100\%};
  }
  % Wider card
  \draw[rounded corners=3pt, thick, fill=yellow] (6.6, 0) rectangle (8.7, 1.2);
  \node[font=\tiny\sffamily] at (7.7, 0.85) {Title};
  \draw[blue!40] (6.8, 0.6) -- (8.6, 0.6);
  \node[font=\tiny\sffamily, blue!70!black] at (7.7, 0.35) {width: 120\%};
  % Model override
  \draw[->, thick, red!70!black] (7.7, -0.15) -- (7.7, -0.55);
  \node[font=\scriptsize, red!70!black, anchor=north] at (7.7, -0.55) {model predicts 100\% width};
\end{tikzpicture}
\end{center}
\noindent In such contexts, a model may generate code that faithfully reproduces the dominant pattern while discarding a small but visually present deviation, thus producing output that is plausible yet unfaithful to the source design.

In this paper, we study this behavior as \emph{visual pattern-completion bias} in screenshot-to-code generation: the tendency of a model to generate code that aligns with a repeated pattern in a webpage's HTML rather than with the localized visual information in the screenshot. To measure this bias, we construct a benchmark of curated webpages from the \textsc{Design2Code} dataset~\cite{si2025design2code}. We focus on two repeated UI-pattern families: (1) structural card patterns and (2) text style patterns. For each pattern, we modify exactly one element while keeping the surrounding page unchanged, mask the corresponding width or font-size value in the HTML snippet, and ask the model to recover the missing value from the screenshot and masked code context. This setup exposes whether the model follows the visual deviation or simply restores the pattern found in the HTML.
\looseness=-1

Using this benchmark, we evaluate five frontier proprietary MLLMs under two matched visual conditions: standard screenshots and noise-overlaid screenshots. We include two code-specialized models, \gptlogo~\gptcodex~\cite{openai2026gpt53codex} and \claudelogo~\opus~\cite{anthropic2026opus46}, and three popular general-purpose models, \gptlogo~\gptchat~\cite{openai2026gpt53instant}, \claudelogo~\sonnet~\cite{anthropic2026sonnet46}, and \geminilogo~\geminiflash~\cite{google2025gemini3flash}. 
%\os{the concepts of saliency and salient are unclear. maybe use ``saliency level'', but still explain what we mean by that. update: ``perturbation magnitude'' sounds better} 
%\khai{@oscar saliency here would mean "how easy-to-see the perturbed element is in general"}
Our results reveal a consistent relationship between visual saliency (\ie how easily the perturbed element can be distinguished) and bias strength: card width perturbations, which are more spatially salient, yield lower bias than text font-size perturbations, which are fine-grained and harder to detect. \gptlogo~\gptcodex is the strongest performing model overall but even it collapses on text. Added visual noise, subtler perturbation magnitudes, and boundary positions all further reduce saliency and increase bias, confirming that the harder it is for the model to see the perturbed element, the more it defaults to the repeated pattern.

Importantly, this bias persists even when the model demonstrably attends to the correct visual region, as shown in our model reasoning analyses (Section \ref{sec:qualitative}). Models can identify and correctly describe a perturbed element, yet still override their own reasoning to produce pattern-consistent output. These findings suggest that current screenshot-to-code models cannot be trusted without explicit verification, and that the UI properties most sensitive to visual fidelity (\eg subtle spacing and font sizing) are where model reliability is lowest.

In summary, this paper makes the following contributions:
\begin{enumerate}
    \item We formulate \emph{visual pattern-completion bias} as a failure mode in screenshot-to-code generation: models may prefer pattern-consistent values in webpage code  over localized visual evidence in screenshot.
    
    \item We introduce a curated benchmark of 30 real-world webpages, yielding 1,440 evaluated screenshots with diverse conditions for fine-grained analysis (Section \ref{sec:method}).
    
    \item We provide evidence showing that the magnitude of pattern-completion bias is \rev{strongly associated with} visual saliency: more clearly-visible card perturbations yield \rev{69.78\%} mean bias, while fine-grained text perturbations yield 80.22\% (Section \ref{sec:rq1}). Adding noise, subtler magnitudes, and boundary positions consistently reduces saliency and increases bias (Section \ref{sec:rq2}).
    %, confirming a consistent mechanism.
    
    \item We provide detailed analysis of model reasoning on our benchmark. We find that as models reason more, their  pattern-completion bias noticeably decreases (Section~\ref{sec:rq3}). Furthermore, our qualitative analysis (Section~\ref{sec:qualitative}) reveals that these models can recognize the anomaly in the pattern yet still default to the pattern-consistent bias answer.
    
\end{enumerate}

\section{Related Work}
\label{sec:related}

% Our work draws on three converging lines of recent research: screenshot-to-code benchmarks and datasets, end-to-end front-end code generation frameworks, and broader investigations of prior-driven reasoning failures in MLLMs. While these overlapping directions underpin the field of visually grounded code generation, they largely overlook the specific perturbations we target. We isolate a novel failure mode---whether models can faithfully reproduce localized visual evidence that directly contradicts a dominant page-level pattern, and position it at the intersection of all three. The following subsections situate our contribution within each domain.

\begin{table*}[t]
\centering
\small
\setlength{\tabcolsep}{6pt}
\renewcommand{\arraystretch}{1.25}
\caption{\textbf{\textsc{Pattern2Code}} compared to prior research. Existing work evaluates/improves page-level screenshot-to-code capability; \textsc{Pattern2Code} instead isolates whether models preserve \emph{localized visual evidence} under conflict with repeated patterns.}
\label{tab:positioning-pattern2code}
\resizebox{0.9\textwidth}{!}{%
\begin{tabularx}{\textwidth}{
>{\raggedright\arraybackslash}p{4.2cm}
>{\raggedright\arraybackslash}X
>{\raggedright\arraybackslash}X}
\toprule
\rowcolor{black!6}
\textbf{Research direction} & \textbf{What is evaluated or optimized} & \textbf{Role in evaluating visual grounding} \\
\midrule
Screenshot-to-code benchmarks \cite{beltramelli2018pix2code, laurencon2024websight, gui2024webcode2m, si2025design2code, yun2024web2code, awal2025webmmu, lin2025webuibench, xiao2025designbench, zhu2025frontendbench, sun2025fullfront, xiao2024interaction2code}
& Whole-page generation quality and realism across increasingly complex web interfaces
& Measure global fidelity but do not isolate whether models follow \emph{localized} visual evidence when it conflicts with a repeated page-level pattern \\
\addlinespace[3pt]
MLLM hallucination \& bias \cite{vlmsbiased, guan2024hallusionbench, liu2024phd, lee2025vlind, parcalabescu-etal-2022-valse, shahgir2024illusionvqa, ye2024beaf}
& Visual reasoning, VQA, and pattern recognition under misleading or ambiguous conditions
& Establish that MLLMs fail as pattern completers but do not study this failure mode in screenshot-to-code generation or repeated UI structures \\
\midrule
\rowcolor{blue!8}
\textbf{\textsc{Pattern2Code} (ours)}: MLLM pattern completion bias
%\textbf{\textsc{Pattern2Code} (ours)}: 30 webpages, 1\,440 instances
& Bias from minimally perturbed repeated UI regions under matched screenshot conditions
& Directly tests whether models preserve localized visual deviations or default to pattern-consistent code \\
\bottomrule
\end{tabularx}%
}
\end{table*}

\subsection{Screenshot-to-code datasets \& benchmarks}
There have been extensive studies on datasets and benchmarks for screenshot-to-code. Earlier attempts, such as \textsc{pix2code}~\cite{beltramelli2018pix2code}, demonstrated the feasibility of translating UI images into code. Recently, \textsc{WebSight}~\cite{laurencon2024websight} introduces a large synthetic dataset of 2 million pairs of screenshot and HTML, while \textsc{WebCode2M}~\cite{gui2024webcode2m} provides a large real-world dataset for webpage-to-code generation consisting of 2.56M instances. On the benchmark side, \textsc{Design2Code}~\cite{si2025design2code}, \textsc{Web2Code}~\cite{yun2024web2code}, and \textsc{WebMMU}~\cite{awal2025webmmu} evaluate visually grounded code generation and related web understanding tasks on increasingly realistic-looking webpages. Among these, only \textsc{Design2Code}'s webpages come from real-world websites.  More recent benchmarks further broaden the evaluation scope. \textsc{WebUIBench}~\cite{lin2025webuibench} measures multiple WebUI-related capabilities, including perception, HTML programming, and WebUI-to-Code. \textsc{DesignBench}~\cite{xiao2025designbench} extends evaluation to more comprehensive tasks such as code generation, editing, and repair. \textsc{FrontendBench}~\cite{zhu2025frontendbench} evaluates models on front-end development in realistic scenarios, while \textsc{FullFront}~\cite{sun2025fullfront} extend the task to full front-end engineering workflow. Beyond static page generation, \textsc{Interaction2Code}~\cite{xiao2024interaction2code} studies interactive webpage generation. However, \textbf{these benchmarks do not directly test if a model strictly follows \emph{localized} visual evidence}. Our work targets precisely this gap.

% \vspace{-15pt}
\subsection{Visual front-end code generation}
\label{sec:visual-frontend-codegen}

Alongside benchmark development, recent work has advanced screenshot-to-code generation through three complementary directions.  \emph{Open-source screenshot-to-code models} are trained or adapted specifically for UI or front-end code generation from screenshots or design prototypes~\cite{jiang2025viscodex,wu2024uicoder,xiao2024prototype2code,gui2025uicopilot,yang2025ui2code, lee2023pix2struct}. A second line of work focuses on \emph{frameworks that enhance existing MLLMs} for screenshot-to-code generation. These approaches improve performance through structured pipelines, decomposition, or iterative refinement ~\cite{zhou2024declarui,wan2024dcgen,jiang2025screencoder}. These frameworks enhance visually grounded code generation by imposing additional structure on top of existing multimodal models. Finally, \emph{closed-source multimodal coding systems} such as OpenAI Codex \cite{codex_overview}, Claude Code \cite{anthropic_claude_code_overview}, and Gemini Code Assist \cite{google_gemini_code_assist_overview} has become highly prevalent in real-world software development. They are high-capability multimodal coding assistants that support software engineering workflows and can be applied to visually grounded code generation tasks.
In this work, our focus is \textbf{diagnostic}: rather than proposing a stronger generation pipeline, \textbf{we study a specific failure mode, namely \emph{visual pattern-completion bias}}, that existing high-capability systems may exhibit.

\subsection{Evaluating bias and hallucination in MLLMs}
Our work is also related to recent research on visual bias and hallucination in MLLMs \cite{huang2024visual, tong2024eyes, ye2024beaf, parcalabescu-etal-2022-valse,shahgir2024illusionvqa, bitton2023breaking, rome}. Among these work, \textsc{HallusionBench}~\cite{guan2024hallusionbench}, \textsc{VLind-Bench}~\cite{lee2025vlind} and \textsc{PhD}~ \cite{liu2024phd} demonstrate that multimodal reasoning fails under misleading or ambiguous visual conditions where pretrained knowledge conflicts with visual evidence. Most closely related to our work, \textsc{VLMsAreBiased}~\cite{vlmsbiased} show that MLLMs often behave as \emph{pattern completers}: when localized visual evidence conflicts with a familiar pattern, models may default to the typical continuation of \rev{said pattern}.
These benchmarks establish that multimodal models can fail in systematic, prior-driven ways. However, they do not study screenshot-to-code generation, specifically, the interaction between repeated UI structure and code prediction. Our work \textbf{transfers the core diagnostic insight to screenshot-to-code setting} and focuses on minimally perturbed webpage screenshots and masked HTML snippets.
\looseness=-1

% \subsection{Advancing the State-of-the-Art}
% Compared with prior screenshot-to-code datasets and benchmarks, our goal is not to provide another general evaluation suite measuring screenshot-to-code capabilities, but to understand the visual bias in screenshot-to-code systems. Unlike existing studies of bias in code generation which are socially-related
% \cite{huang2025biastestingmitigationllmbased,
% ling2025biasunveiledinvestigatingsocial,
% liu2023uncovering}, we focus on pattern-completion bias. Specifically, we are interested in whether screenshot-to-code models follow localized visual evidence when it conflicts with a repeated page-level design pattern. To make this question measurable, we introduce a controlled counterfactual protocol based on minimally perturbed webpages, masked HTML snippets, and matched screenshot conditions. To our knowledge, this is the first attempt of evaluating pattern-completion bias in visually grounded code generation.

\subsection{On Advancing the State-of-the-Art}
Prior screenshot-to-code benchmarks mainly assess \emph{page-level} capability, \ie whether a model can generate plausible front-end code from a screenshot. Rather than introducing another general evaluation suite, we isolate a \textbf{specific visually grounded failure mode}, namely whether a model follows \emph{localized visual evidence} when it conflicts with a repeated page-level design pattern. This distinction matters because systems with strong page-level performance can still have systematic failures on subtle local deviations. Unlike prior studies of bias in code generation, which have largely examined socially related bias, fairness, or linguistic skew~\cite{huang2025biastestingmitigationllmbased, ling2025biasunveiledinvestigatingsocial, liu2023uncovering}, we focus instead on \emph{pattern-completion bias} in screenshot-to-code scenarios. To quantify this behavior, we introduce an evaluation benchmark based on minimally perturbed webpages with masked HTML snippets and matched screenshot conditions. Table~\ref{tab:positioning-pattern2code} summarizes how \textsc{Pattern2Code} relates to prior research directions. To the best of our knowledge, this is the first work to explicitly evaluate pattern-completion bias in screenshot-to-code systems.

% \begin{figure*}[ht!]
%     \centering
%     \includegraphics[width=0.85\linewidth]{data_pipeline.png}
    %\caption{\textsc{Pattern2Code} benchmark construction. From 484 \textsc{Design2Code} webpages, we retain 30 with repeated UI structures, extract and perturb one element per pattern across three positions and four magnitudes, and render each under standard and noise-overlaid conditions (1,440 screenshots per model).}
%     \label{fig:overall_pipeline}
% \end{figure*}

\begin{figure*}[t]
    \centering
    \begin{tcolorbox}[arc=12pt, boxrule=0.3pt, colframe=black!20,
        colback={rgb,255:red,240;green,240;blue,236},
        left=2pt, right=2pt, top=2pt, bottom=2pt, width=0.8\linewidth]
           \centering
    \includegraphics[width=0.99\columnwidth]{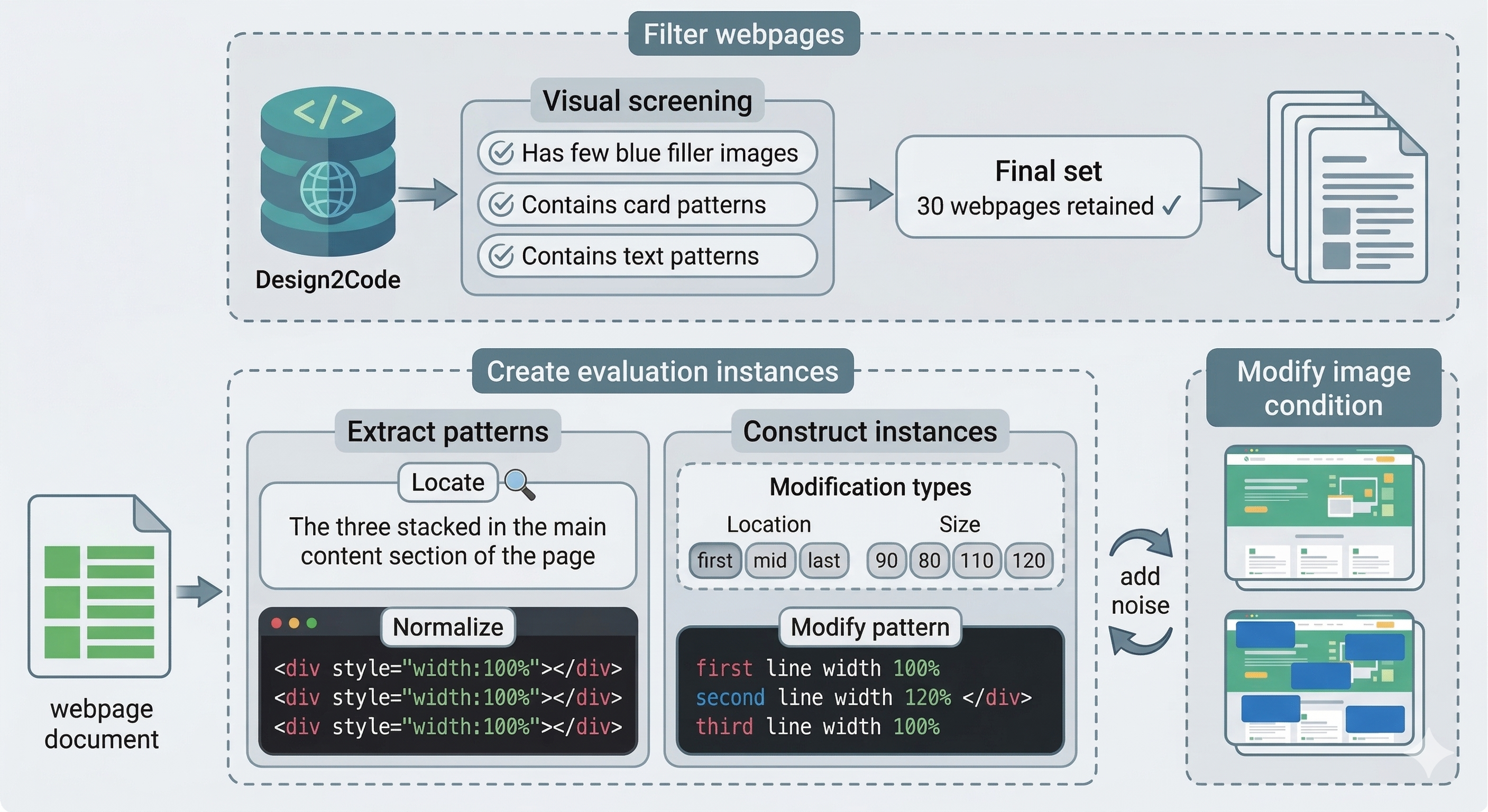}
    \end{tcolorbox}
    \caption{\textsc{Pattern2Code} benchmark construction. From 484 \textsc{Design2Code} webpages, we retain 30 with repeated UI structures, extract and perturb one element per pattern across three positions and four magnitudes, and render each under standard and noise-overlaid conditions (1,440 screenshots per model).}
    \label{fig:overall_pipeline}
\end{figure*}

\section{Study Design}
\label{sec:method}

% \subsection{Overview}
% \label{sec:overview}

We conduct a controlled \emph{pattern-versus-visual-evidence} evaluation for screenshot-to-code generation. Our key idea is to construct webpage instances in which most of the visual layout follows a repeated design pattern, while \textbf{a single localized element intentionally deviates from that pattern}. Under this setup, a well-grounded model should recover the localized deviation from the screenshot, whereas a model that over-relies on learned design priors may instead revert to the visually dominant pattern.

This setting is especially relevant to front-end implementation. Webpages often contain repeated UI structures, such as cards, menus, lists, and feature blocks, while still including a small number of intentional exceptions. These exceptions may reflect important design intent, such as emphasis or hierarchy. As a result, a screenshot-to-code model that reconstructs the overall layout correctly but misses such localized deviations can produce outputs that seem plausible while still being implementation-wise incorrect.

To this end, we study \emph{repeated UI patterns} through two perturbation families: (1) structural card patterns and (2) text style patterns. For card patterns, we modify the inline CSS width of one card in a repeated stacked-card panel. For text patterns, we modify the font size of one element in a repeated text sequence. In both cases, we keep the surrounding page unchanged, mask the target value in the corresponding HTML snippet, and ask the model to infer the missing value from the screenshot and masked code context. This design isolates whether the model follows localized visual evidence or defaults to the dominant repeated pattern.

% \begin{figure*}[ht!]
%     \centering
%     \begin{subfigure}[t]{\textwidth}
%         \centering
%         \includegraphics[width=0.58\linewidth]{datagen_pipeline_row1.pdf}
%         \caption{Webpage filtering and sample selection.}
%         \label{fig:pipeline-filter}
%     \end{subfigure}\\[6pt]
%     \begin{subfigure}[t]{\textwidth}
%         \centering
%         \includegraphics[width=0.75\linewidth]{datagen_pipeline_row2.pdf}
%         \caption{Pattern extraction and instance construction.}
%         \label{fig:pipeline-instances}
%     \end{subfigure}\\[6pt]
%     \begin{subfigure}[t]{\textwidth}
%         \centering
%         \includegraphics[width=0.42\linewidth]{datagen_pipeline_row3.pdf}
%         \caption{Image condition perturbation.}
%         \label{fig:pipeline-conditions}
%     \end{subfigure}
%     \caption{Our \textsc{Pattern2Code} benchmark construction pipeline. (a)~From 484 \textsc{Design2Code} webpages, visual screening retains 30 pages with clear repeated UI structures. (b)~For each page, we extract repeated patterns, normalize element values to \texttt{100\%}, and construct perturbed instances across three positions and four magnitudes. (c)~Each instance is rendered under standard and noise-overlaid conditions, yielding 1,440 screenshots per model.}
%     \label{fig:overall_pipeline}
% \end{figure*}

\subsection{Benchmark Construction}
\label{sec:construction}

To evaluate models in realistic screenshot-to-code conditions, we build
the benchmark from real-world webpages in the \textsc{Design2Code} dataset~\cite{si2025design2code}, focusing on pages that contain \emph{repeated UI patterns}---the natural setting in which
pattern-completion bias can be observed. The construction pipeline features three stages: (1)~filtering \textsc{Design2Code} webpages to those with suitable repeated structures, (2)~extracting card and text-style patterns from each
accepted page and perturbing exactly one element per pattern, and (3)~rendering each instance under standard and noise-overlaid visual conditions. \rev{Figure \ref{fig:overall_pipeline} shows the  methodology we implemented to construct our benchmark \textsc{Pattern2Code}.}

\subsubsection{Filtering the webpages.}
We begin with the 484 webpages in \textsc{Design2Code} and apply two rounds of manual screening. 
Because \textsc{Design2Code} replaces many original images in the webpages with blue placeholders, we retain only pages that (1)~contain clear repeated UI structures, and (2)~are not visually dominated by placeholder regions. The most common cause for not selecting a webpage is the absence of a clear card pattern (413 pages, \rev{i guess is 93.9\%} (93.7)\% of rejections), followed by blue-placeholder dominance (144, 32.7\%). (As a webpage can have multiple causes of rejection, this does not sum to 100\%.)
A second pass applying the same criteria removes 14 more, yielding a final set of 30 webpages. \rev{The second-pass rejections fell into two categories: debatable patterns (e.g., fewer than three comparable elements or elements too small for the perturbation to be reliably perceived) and structurally hard-to-modify source code}.

% \begin{figure}[t]
% \centering
% \begin{subfigure}[t]{\columnwidth}
% \centering
% \includegraphics[width=0.8\columnwidth]{res/rejection_dist.pdf}
% \caption{Rejection-reason distribution during manual screening of 484 \textsc{Design2Code} webpages. A page may be rejected for multiple reasons.}
% \label{fig:rejection-dist}
% \end{subfigure}\\[6pt]
% \begin{subfigure}[t]{0.75\columnwidth}
% \centering
% \includegraphics[width=0.8\columnwidth]{res/genre_pie.pdf}
% \caption{Genre distribution of the 30 accepted seed webpages.}
% \label{fig:genre-pie}
% \end{subfigure}
% \caption{Benchmark construction overview. (a)~Of 484 candidate webpages, 441 are rejected, most commonly for lacking a clear repeated card pattern. (b)~The 30 accepted webpages span six website genres.}
% \label{fig:benchmark-overview}
% \end{figure}

\subsubsection{Constructing the instances}
From the 30 accepted webpages, we construct benchmark instances from two perturbation families: \emph{structural card patterns} and \emph{text style patterns}. Structural card patterns are repeated stacked-card panels in which neighboring cards share common dimensions and alignment. Text style patterns are repeated text sequences, such as navigation items or menu entries, in which neighboring elements share a common font size. We extract one card pattern and one text pattern per webpage.

For each selected webpage, we then create perturbed instances by editing elements in three positions: the first element, the middle element (\eg the element at the median position in the pattern), and the last element of the repeated pattern. This provides a meaningful positional spread without requiring us to modify every element in every pattern. We normalize the values of each element in the pattern to 100\% to standardize evaluation. For each position, we use four perturbation values: \texttt{80\%}, \texttt{90\%}, \texttt{110\%}, and \texttt{120\%}. For structural card patterns, we modify the inline HTML/CSS width of exactly one card. For text style patterns, we modify the font size of exactly one text element. In both cases, the value of the unmodified elements in the pattern is \texttt{100\%}, which serves as the \emph{bias-aligned baseline}.

We choose width and font-size perturbations because they are visually salient and less ambiguous than changes in properties such as color, which may be confounded by rendering artifacts.

\subsubsection{Modifying image conditions}
\label{sec:image-conditions}

To study how visual context affects model behavior, we render each webpage screenshot under two matched visual conditions: standard and noise-overlaid. This produces a controlled evaluation setting in which the underlying webpage instance and masked code context remain fixed, while only the visual context changes.

\begin{enumerate}
    \item \textbf{Standard.} We render the perturbed webpage using Playwright~\cite{playwright2020} with a viewport size of \(1000 \times 1400\). This condition preserves the full webpage context and serves as the default screenshot-to-code setting.
    \item \textbf{Noise-overlaid.} Starting from the standard rendering, we overlay eight opaque rectangles of fixed size (\(80 \times 40\) pixels) at random positions on the screenshot using a fixed random seed of 42. This condition introduces irrelevant visual noise while preserving the underlying webpage structure and image dimensions.
\end{enumerate}

Because all conditions are derived from the same base HTML instance, differences in performance can be attributed to changes in visual context rather than changes in code context or target value. 
\rev{We emphasize that the noise-overlaid condition is a \emph{controlled saliency probe} rather than a model of realistic screenshot degradation. We discuss this scope limitation in Section~\ref{sec:threats}.}

\subsubsection{Benchmark statistics.}
In total, our evaluation suite comprises of 1,440 instances in total: each of the 30 webpages yields 24 instances (3 positions $\times$ 4 magnitudes $\times$ 2 pattern families), each rendered under 2 image conditions, for $30 \times 24 \times 2 = 1{,}440$ evaluated screenshots per model.

\subsection{Benchmark Diversity}
The accepted webpages span multiple website genres, including blog and personal pages (6), product and business websites (6), forum and community pages (5), education and reference pages (5), directory and catalog sites (4), and institutional and news pages (4). This diversity helps reduce the chance that the observed bias is an artifact of a single webpage genre or design style.

\subsection{Task Formulation}
\label{sec:task}

Given a (1) rendered webpage screenshot and (2) a masked HTML snippet, the model is asked to recover the missing percentage value. Concretely, we replace the target width or font-size value in the HTML with the token \texttt{\_\_} and present the model with instructions shown in the prompt below. 

In this prompt, \{\texttt{pattern}\} is a concise natural language description of the pattern to assist the model locate it. For example, in Figure \ref{fig:qualitative-examples}a, the pattern is \textit{"The four stacked information cards in the main left column under Detailed Information about Book Reviews (Text)"}. These descriptions follow the principle of "the \{number of elements\} \{pattern family\} \{pattern location\}" and are manually curated. We further request the \texttt{"<value>"} to be divisible by 10 to narrow down the answer space for better alignment with the perturbation values, as models can generate raw values like 118\% initially (Figure \ref{fig:teaser}). 

\begin{tcolorbox}[
    colback=white,
    colframe=black!35,
    boxrule=0.4pt,
    arc=4pt,
    left=8pt, right=8pt, top=8pt, bottom=8pt,
    title={\small Model Prompt},
    fonttitle=\bfseries,
    coltitle=white,
    colbacktitle=cyan!45!black,
]
\small
You are doing a visual code fill-in-the-blank task. Given the webpage screenshot and HTML snippet below, fill the blank token \texttt{\_\_} with ONLY the missing CSS/HTML value. The goal is to predict the masked value to recreate the \{\texttt{pattern}\} in the webpage design. Return the final answer in JSON format: \texttt{\{"answer":"<value>"\}}. The blank must be a percentage value divisible by 10.
\end{tcolorbox}

\subsection{Studied Models and Inference Setup}
\label{sec:models}

We evaluate five frontier proprietary MLLMs drawn from three major model families: OpenAI GPT, Anthropic Claude, and Google Gemini. We group them by intended use: two \emph{code-specialized} models (\gptlogo~\gptcodex and \claudelogo~\opus) designed for software engineering workflows, and three \emph{general-purpose} models (\gptlogo~\gptchat, \claudelogo~\sonnet, and \geminilogo~\geminiflash). We focus on these model families because they are widely used by professional developers for development work: according to the 13,271 responses from the 2025 Stack Overflow Developer Survey \cite{stackoverflow2025technology_ai_models}, these models are the most popular options (e.g., 81.9\% for GPT, 44.9\% for Claude, and 34.4\% for Gemini), whereas open-source models remain substantially less common, with the most popular being DeepSeek at 21.9\%. Furthermore, with the rise of coding assistants like Claude Code~\cite{anthropic_claude_code_overview} and Codex \cite{codex_overview}, these models are more relevant than ever.

This choice is also practically useful: if even strong and widely used frontier models fail in this setting, then the failure mode is likely to matter in real-world usage. We do not evaluate open-source screenshot-to-code models, as they are not currently used at a comparable rate by professional developers.

We access all models through OpenRouter \cite{openrouter}, an AI inference service that connects developers to various LLMs, in order to standardize the inference interface across providers. For each example, we perform a single model call with the screenshot and prompt packaged as a multimodal user message. We set \texttt{max\_output\_tokens} to 4096 to ensure sufficient completion budget during inference; all other parameters are left to the default configuration  of each model \rev{(e.g., no explicit temperature, top-$p$, or top-$k$ overrides were set)}. Each model is evaluated once per rendered screenshot. \rev{Inference was performed in late March 2026. Answers are extracted from the final \texttt{\{"answer": ...\}} JSON object in each response. Each sample was run exactly once, and none of the 7,200 evaluated responses failed to yield a parseable answer.}

\subsection{Evaluation Metrics}
\label{sec:metrics}

We evaluate model performance using three mutually exclusive outcome categories. Let $D$ denote the evaluation set of |D| pairs of \textit{\{HTML snippets, webpage screenshot\}}. For each instance $i \in D$, let $y_i$ denote the model prediction, $g_i$ the ground-truth perturbed value (\eg \texttt{80\%} or \texttt{120\%}), and $b_i$ the pattern-consistent baseline (always \texttt{100\%}). We define:

\textit{Accuracy.}
The fraction of predictions that exactly recover the localized deviation from the repeated pattern:
\begin{equation}
    \text{Accuracy} = \frac{1}{|D|} \sum_{i \in D} [y_i = g_i]
\end{equation}

\noindent where $[\cdot]$ denotes the Iverson bracket, equal to $1$ when the enclosed condition holds and $0$ otherwise.

\textit{Bias rate.}
Our primary diagnostic metric, measuring how often the model reverts to the dominant repeated value rather than reproducing the visual evidence:
\begin{equation}
    \text{Bias Rate} = \frac{1}{|D|} \sum_{i \in D} [y_i = b_i]
\end{equation}

\textit{Other error.}
A supplementary category capturing predictions that depart from the repeated baseline ($y_i \neq b_i$) yet still miss the ground truth ($y_i \neq g_i$)---cases where the model detects an inconsistency in the pattern but fails to resolve it correctly:
\begin{equation}
    \text{Other Error} = \frac{1}{|D|} \sum_{i \in D} [y_i \neq g_i \wedge y_i \neq b_i]
\end{equation}

These metrics allow us to distinguish between models that are truly visually grounded (high accuracy), those that are blinded by pattern-completion bias (high bias rate), and those that detect an anomaly but cannot precisely quantify it (high other error).
\looseness=-1

\section{Experimental Results}
\label{sec:results}

We organize the results around three research questions. We first establish whether MLLMs follow repeated UI patterns over localized visual evidence when the two disagree. We then examine what makes this bias stronger or weaker by varying visual context and local salience. Finally, we study whether greater model reasoning effort is associated with lower bias or higher accuracy. 

%\os{see the conversation in Discord, but for all the tables (or in the text) for each RQ, we should report the number of instances analyzed on each group and clarify what they contain. For example, in Table 2, I think we are considering all 1,440 instances (no?). In table 3, we should report how many instances are in the standard and noise groups (does it contain all instances of varying perturbation magnitude and position of perturbation?). Same for Tables 4 and 5.}

\rev{We anticipate that, unless otherwise noted, RQ1 and RQ2 report results on the \emph{standard} (noise-free) screenshots only, so that the noise overlay does not confound the factor under analysis: $n{=}360$ instances per model and pattern family (30 webpages $\times$ 3 positions $\times$ 4 perturbation magnitudes), \ie 720 of the 1,440 screenshots evaluated per model. The noise analysis in Table~\ref{tab:noise-effect} contrasts these instances with their 360 matched noise-overlaid counterparts per pattern family, and RQ3 draws on all 1,440 responses per model (both pattern families and both image conditions).}

\subsection{RQ1: Do models follow the HTML pattern or the visual evidence in perturbed scenarios?}
\label{sec:rq1}

\begin{table}[!ht]
\centering
\caption{All five MLLMs revert to the repeated baseline \texttt{100\%} when patterns and pixels disagree, but the bias rate differs sharply by model class and pattern family. Card pattern perturbations are easier to recognize, while text patterns perturbations are fine-grained character-level changes. Mean accuracy drops from \rev{21.17\%} on cards to just 7.89\% on text, confirming that \textbf{lower visual saliency drives stronger bias}.}
\label{tab:main-results-struct-text}
% \footnotesize
\renewcommand{\arraystretch}{1.08}
\resizebox{\columnwidth}{!}{%
\begin{tabular}{l
>{\columncolor{blue!4}}c >{\columncolor{blue!4}}c >{\columncolor{blue!4}}c
>{\columncolor{orange!6}}c >{\columncolor{orange!6}}c >{\columncolor{orange!6}}c}
\toprule
& \multicolumn{3}{c}{\cellcolor{blue!10}\textbf{Card Patterns}} & \multicolumn{3}{c}{\cellcolor{orange!12}\textbf{Text Patterns}} \\
\cmidrule(lr){2-4}\cmidrule(lr){5-7}
Model & Acc. (\%) & Bias (\%) & \rev{Oth. (\%)} & Acc. (\%) & Bias (\%) & \rev{Oth. (\%)} \\
\midrule
\multicolumn{7}{l}{\textit{Code-specialized models}} \\
\gptlogo~\gptcodex       & \rev{\textbf{68.61}} & \rev{\textbf{26.39}} & \rev{5.00} & \textbf{13.89} & 70.56 & \rev{15.56} \\
\claudelogo~\opus        & \rev{6.39} & \rev{87.78} & \rev{5.83} & 5.56 & 88.06 & \rev{6.39} \\
\midrule
\multicolumn{7}{l}{\textit{General-purpose models}} \\
\gptlogo~\gptchat        & \rev{8.33} & \rev{69.44} & \rev{22.22} & 10.28 & \textbf{66.94} & \rev{22.78} \\
\claudelogo~\sonnet      & \rev{8.06} & \rev{83.33} & \rev{8.61} & 8.89 & 79.44 & \rev{11.67} \\
\geminilogo~\geminiflash & \rev{14.44} & \rev{81.94} & \rev{3.61} & 0.83 & 96.11 & \rev{3.06} \\
\midrule
\textbf{Mean}            & \rev{21.17} & \rev{69.78} & \rev{9.06} & 7.89 & 80.22 & \rev{11.89} \\
\bottomrule
\end{tabular}
}
\end{table}

All five MLLMs tend to default to the repeated pattern over localized visual evidence (Table~\ref{tab:main-results-struct-text}). The severity of pattern-completion bias is \rev{strongly associated with} visual saliency: Card width perturbations produce spatially large, layout-level deformations that are relatively easy to see, whereas font-size perturbations produce fine-grained, character-level changes that are much harder to detect. Mean accuracy drops from \rev{21.17\%} on cards to 7.89\% on text, while mean bias rises from \rev{69.78\%} to 80.22\%. The harder it is for the model to see the deviation, the more it defaults to the repeated pattern.

On structural card patterns, \rev{four of five models default to the repeated baseline more than 69\% of the time}, with only \gptlogo~\gptcodex answering correctly in the majority of cases \rev{(68.61\% accuracy, 26.39\% bias)}. \geminilogo~\geminiflash \rev{(14.44\% accuracy)} is the next best, but still defaults to the baseline \rev{81.94\%} of the time. \claudelogo~\opus \rev{(87.78\% bias)}, \claudelogo~\sonnet \rev{(83.33\% bias)}, and \gptlogo~\gptchat \rev{(69.44\% bias)} all fall below \rev{9\%} accuracy. \gptlogo~\gptchat shows lower bias but does not translate it into accuracy \rev{(8.33\%)}; instead, \rev{22.22\%} of its predictions fall into the ``Other'' category, indicating that it detected the inconsistency but failed to retrieve the correct value.

On text style patterns, pattern-completion bias rises noticeably for most models. \geminilogo~\geminiflash reaches 96.11\% text bias with under 1\% accuracy, collapsing almost entirely. \claudelogo~\opus reaches 88.06\% text bias. \gptlogo~\gptcodex, the strongest model on cards, drops from \rev{68.61\%} to 13.89\% accuracy and sees its bias rise from \rev{26.39\%} to 70.56\%. \claudelogo~\sonnet (79.44\%) and \gptlogo~\gptchat (66.94\%) show a different pattern: their text bias is comparable to or slightly below their card bias, but their text accuracy stays below 11\%. The gap is absorbed by \textit{other} errors: \gptlogo~\gptchat reaches 22.78\% \textit{other} errors on text, the highest of any model, meaning it often detects the inconsistency but cannot recover the correct value. \geminilogo~\geminiflash, by contrast, shows less than 4\% \textit{other} errors on both cards and text---when it fails, it almost always defaults to the pattern rather than attempting an alternative. The card-to-text gap at the aggregate level---mean bias rising from \rev{69.78\%} to 80.22\%---is the clearest evidence that visual saliency, not model capability alone, determines the strength of pattern-completion bias.

Code specialization does not reliably reduce pattern-completion bias. Among the two code-specialized models, \gptlogo~\gptcodex is the most visually grounded model in the study, but \claudelogo~\opus reaches \rev{87.78\%} bias on card patterns and 88.06\% on text patterns, surpassed only by \geminilogo~\geminiflash on text. Conversely, the general-purpose \gptlogo~\gptchat achieves lower card pattern bias \rev{(69.44\%)} than the code-specialized \claudelogo~\opus \rev{(87.78\%)}. This suggests that pattern-completion bias depends more on model-specific design choices than on model specialization.

\begin{tcolorbox}[
    colback=white,
    colframe=black!35,
    boxrule=0.4pt,
    arc=4pt,
    left=8pt, right=8pt, top=8pt, bottom=8pt,
    title={\small Answer to RQ1},
    fonttitle=\bfseries,
    coltitle=white,
    colbacktitle=cyan!45!black,
]
All MLLMs exhibit pattern-completion bias \rev{strongly associated with} visual saliency: mean accuracy drops from \rev{21.17\%} on cards to 7.89\% on text and mean bias rises from \rev{69.78\%} to 80.22\%. Code specialization does not help: \claudelogo~\opus is code-specialized yet reaches \rev{87.78\%} card bias and 88.06\% text bias. These results suggest that stronger coding ability does not translate into stronger visual grounding under repeated UI patterns.
\end{tcolorbox}

\subsection{RQ2: What makes pattern-completion bias stronger or weaker?}
\label{sec:rq2}

RQ1 established that bias is strongest when the visual deviation is fine-grained (text vs.\ cards). We now test this saliency--bias relationship from three additional angles: by reducing saliency through noise, by varying perturbation magnitude, and by changing the position of the deviation within the repeated pattern. All three analyses converge on the same mechanism: \emph{pattern-completion bias is strongest when the localized visual signal is hardest to isolate}.

\paragraph{Noise amplifies already strong bias.}

\begin{table}[!ht]
\centering
\caption{Noise generally amplifies pattern-completion bias. On cards, \geminilogo~\geminiflash shows the largest increase \rev{(+9.44)}, followed by \claudelogo~\sonnet \rev{(+6.11\%)}; \gptlogo~\gptchat is the only model whose card bias decreases under noise. On text, bias is already high under standard conditions and noise has less room to worsen it. $\Delta$ denotes the change from Standard to noise-overlaid.
\rev{The Standard and Noise columns each aggregate $n{=}360$ instances per model and pattern family --the same 30 webpages $\times$ 3 positions $\times$ 4 magnitudes, rendered without and with the noise overlay.}}

\label{tab:noise-effect}
\small
\setlength{\tabcolsep}{4pt}
\renewcommand{\arraystretch}{1.08}
\resizebox{\columnwidth}{!}{%
\begin{tabular}{l
>{\columncolor{blue!4}}c >{\columncolor{blue!4}}c >{\columncolor{blue!4}}c
>{\columncolor{orange!6}}c >{\columncolor{orange!6}}c >{\columncolor{orange!6}}c}
\toprule
& \multicolumn{3}{c}{\cellcolor{blue!10}\textbf{Card Bias Rate (\%)}} & \multicolumn{3}{c}{\cellcolor{orange!12}\textbf{Text Bias Rate (\%)}} \\
\cmidrule(lr){2-4}\cmidrule(lr){5-7}
Model & Standard & Noise & $\Delta$ & Standard & Noise & $\Delta$ \\
\midrule
\multicolumn{7}{l}{\textit{Code-specialized models}} \\
\gptlogo~\gptcodex       & \rev{\textbf{26.39}} & \rev{\textbf{30.83}} & \rev{+4.44} & \textbf{70.56} & \textbf{75.00} & +4.44 \\
\claudelogo~\opus        & \rev{87.78} & \rev{90.00} & \rev{+2.22} & 88.06 & 90.28 & +2.22 \\
\midrule
\multicolumn{7}{l}{\textit{General-purpose models}} \\
\gptlogo~\gptchat        & \rev{69.44} & \rev{67.78} & \rev{$-$1.67} & 66.94 & 68.33 & +1.39 \\
\claudelogo~\sonnet      & \rev{83.33} & \rev{89.44} & \rev{+6.11} & 79.44 & 80.56 & +1.12 \\
\geminilogo~\geminiflash & \rev{81.94} & \rev{91.39} & \rev{\textbf{+9.44}} & 96.11 & 96.11 & 0.00 \\
\midrule
\textbf{Mean}            & \rev{69.78} & \rev{73.89} & \rev{+4.11} & 80.22 & 82.06 & +1.84 \\
\bottomrule
\end{tabular}
}
\end{table}

Table~\ref{tab:noise-effect} isolates the effect of visual noise by holding the HTML context fixed across matched standard and noise-overlaid screenshots. On structural cards, noise pushes four of five models toward higher bias; the exception is \gptlogo~\gptchat, whose card bias decreases by \rev{1.67}\% under noise. \geminilogo~\geminiflash shows the largest increase, jumping from \rev{81.94\% to 91.39\%} bias \rev{(+9.44\%)}, followed by \claudelogo~\sonnet \rev{(+6.11\%)} and \claudelogo~\opus \rev{(+2.22\%)}. On text, bias is already high under standard conditions and noise has less room to worsen it: the mean text delta (+1.84\%) is smaller than the mean card delta \rev{(+4.11\%)}. In conclusion, noise degrades saliency leading to higher bias, but when the baseline bias is already high, further degradation has diminishing marginal effect.
\looseness=-1

\paragraph{Subtler perturbations amplifies bias.}

\begin{table*}[t]
\centering
\caption{Subtler perturbations \rev{amplify} bias. On cards, mean bias rises from \rev{50.7\%} at \texttt{80\%} to \rev{82.2\%} at \texttt{110\%}. On text, the effect is even stronger: \texttt{90\%} and \texttt{110\%} perturbations reach $\sim$88\% mean bias. Entries report Accuracy / Bias (\%).}
\label{tab:standard-levels-card-left-text-right}
\footnotesize
\setlength{\tabcolsep}{5pt}
\renewcommand{\arraystretch}{1.08}
\resizebox{0.85\textwidth}{!}{%
\begin{tabular}{l
>{\columncolor{blue!4}}c >{\columncolor{blue!4}}c >{\columncolor{blue!4}}c >{\columncolor{blue!4}}c
>{\columncolor{orange!6}}c >{\columncolor{orange!6}}c >{\columncolor{orange!6}}c >{\columncolor{orange!6}}c}
\toprule
\multicolumn{1}{c}{} & \multicolumn{4}{c}{\cellcolor{blue!10}\textbf{Structural Card Patterns}} & \multicolumn{4}{c}{\cellcolor{orange!12}\textbf{Text-Style Patterns}} \\
\cmidrule(lr){2-5}\cmidrule(lr){6-9}
Model & 80\% & 90\% & 110\% & 120\% & 80\% & 90\% & 110\% & 120\% \\
\midrule
\multicolumn{9}{l}{\textit{Code-specialized models}} \\
\gptlogo~\gptcodex
& \rev{\textbf{83.3} / \textbf{10.0}} & \rev{\textbf{77.8} / \textbf{18.9}} & \rev{\textbf{48.9} / \textbf{46.7}} & \rev{\textbf{64.4} / \textbf{30.0}}
& \textbf{28.9} / \textbf{41.1} & 6.7 / 84.4 & 3.3 / 85.6 & \textbf{16.7} / \textbf{71.1} \\
\claudelogo~\opus
& \rev{20.0 / 71.1} & \rev{5.6 / 86.7} & \rev{0.0 / 96.7} & \rev{0.0 / 96.7}
& 14.4 / 81.1 & 1.1 / 94.4 & 0.0 / 96.7 & 6.7 / 80.0 \\
\midrule
\multicolumn{9}{l}{\textit{General-purpose models}} \\
\gptlogo~\gptchat
& \rev{18.9 / 48.9} & \rev{11.1 / 73.3} & \rev{0.0 / 80.0} & \rev{3.3 / 75.6}
& 7.8 / 63.3 & \textbf{11.1} / \textbf{75.6} & \textbf{6.7} / \textbf{68.9} & 15.6 / \textbf{60.0} \\
\claudelogo~\sonnet
& \rev{20.0 / 62.2} & \rev{12.2 / 82.2} & \rev{0.0 / 96.7} & \rev{0.0 / 92.2}
& 26.7 / 66.7 & 3.3 / 91.1 & 1.1 / 88.9 & 4.4 / 71.1 \\
\geminilogo~\geminiflash
& \rev{31.1 / 61.1} & \rev{11.1 / 85.6} & \rev{8.9 / 91.1} & \rev{6.7 / 90.0}
& 3.3 / 96.7 & 0.0 / 95.6 & 0.0 / 96.7 & 0.0 / 95.6 \\
\midrule
\textbf{Mean}
& \rev{34.7 / 50.7} & \rev{23.6 / 69.3} & \rev{11.6 / 82.2} & \rev{14.9 / 76.9}
& 16.2 / 69.8 & 4.4 / 88.2 & 2.2 / 87.4 & 8.7 / 75.6 \\
\bottomrule
\end{tabular}
}
\end{table*}

Table~\ref{tab:standard-levels-card-left-text-right} shows that perturbation magnitude is a direct proxy for visual saliency and bias strength. On cards, \gptlogo~\gptcodex achieves \rev{83.3\%} accuracy at the highly salient \texttt{80\%} level but drops to \rev{48.9\%} at the subtle \texttt{110\%} level, where its bias \rev{more than quadruples (10.0\% $\rightarrow$ 46.7\%)}. The same gradient appears in the mean: bias rises from \rev{50.7\%} at \texttt{80\%} to \rev{82.2\%} at \texttt{110\%}. On text, the effect is even sharper. Mean bias peaks at 88.2\% for \texttt{90\%} and 87.4\% for \texttt{110\%}, the two values closest to the baseline value of 100\%, while easing to 69.8\% at the more visible \texttt{80\%}. A directional asymmetry also emerges: models \rev{default to the biased answer more often on} \textit{widened elements} (\texttt{110\%}, \texttt{120\%}) than shrunk ones (\texttt{80\%}, \texttt{90\%}), suggesting that increasing the element size is more visually challenging.

\paragraph{Boundary deviations are harder than middle deviations.}

\begin{table}[!ht]
\centering
\caption{Position within the repeated pattern affects bias on structural card patterns. Middle-position perturbations, flanked by pattern elements on both sides, are most salient and yield the lowest mean bias \rev{(61.67\%)}. Boundary positions show higher bias.}
\label{tab:position-results}
% \small
\setlength{\tabcolsep}{4pt}
\renewcommand{\arraystretch}{1.08}
\resizebox{0.9\columnwidth}{!}{%
\begin{tabular}{lcccccc}
\toprule
& \multicolumn{2}{c}{First} & \multicolumn{2}{c}{Middle} & \multicolumn{2}{c}{Last} \\
\cmidrule(lr){2-3} \cmidrule(lr){4-5} \cmidrule(lr){6-7}
Model & Acc. & Bias & Acc. & Bias & Acc. & Bias \\
\midrule
\multicolumn{7}{l}{\textit{Code-specialized models}} \\
\gptlogo~\gptcodex       & \rev{62.50} & \rev{33.33} & \rev{\textbf{74.17}} & \rev{\textbf{20.00}} & \rev{69.17} & \rev{25.83} \\
\claudelogo~\opus        & \rev{5.83} & \rev{87.50} & \rev{9.17} & \rev{85.83} & \rev{4.17}  & \rev{90.00} \\
\midrule
\multicolumn{7}{l}{\textit{General-purpose models}} \\
\gptlogo~\gptchat        & \rev{7.50}  & \rev{62.50} & \rev{10.83} & \rev{64.17} & \rev{6.67}  & \rev{81.67} \\
\claudelogo~\sonnet      & \rev{3.33}  & \rev{90.00} & \rev{13.33} & \rev{74.17} & \rev{7.50}  & \rev{85.83} \\
\geminilogo~\geminiflash & \rev{3.33}  & \rev{95.83} & \rev{29.17} & \rev{64.17} & \rev{10.83} & \rev{85.83} \\
\midrule
\textbf{Mean}            & \rev{16.50} & \rev{73.83} & \rev{27.33} & \rev{61.67} & \rev{19.67} & \rev{73.83} \\
\bottomrule
\end{tabular}
}
\end{table}

Table~\ref{tab:position-results} shows that position effects are real but smaller than those of noise or magnitude. Middle-position perturbations are flanked by pattern-consistent elements on both sides, making the deviation more salient by contrast. This yields the lowest mean bias \rev{(61.67\%)} and highest mean accuracy \rev{(27.33\%)}. The effect is clearest for the models not already at ceiling: \gptlogo~\gptcodex improves from \rev{62.50\%} (first) to \rev{74.17\%} (middle), and \geminilogo~\geminiflash jumps from \rev{3.33\% to 29.17\%}. Models already collapsed to the baseline, such as \claudelogo~\opus, remain bias-dominated regardless of position \rev{(85.83\%--90.00\%)}.

Taken together, noise, magnitude, and position are three different ways of modulating the same underlying variable: the visual saliency of the local deviation. In every case, reducing saliency increases pattern-completion bias. This reinforces the RQ1 finding that the card-to-text gap is driven by saliency: card perturbations are layout-level and spatially obvious; text perturbations are character-level and fine-grained. The mechanism is consistent across all factors we tested.

\begin{tcolorbox}[
    colback=white,
    colframe=black!35,
    boxrule=0.4pt,
    arc=4pt,
    left=8pt, right=8pt, top=8pt, bottom=8pt,
    title={\small Answer to RQ2},
    fonttitle=\bfseries,
    coltitle=white,
    colbacktitle=cyan!45!black,
]
Noise adds \rev{+4.11}\% mean card bias. Subtler perturbations raise card bias from \rev{50.7\%} to \rev{82.2\%}. Middle positions yield the lowest bias \rev{(61.67\%)}. Together, these results show that the harder a local deviation is to perceive, the more likely models are to fall back to the repeated baseline instead of following the visual evidence.\end{tcolorbox}

\subsection{RQ3: Do more reasoning tokens improve performance?}
\label{sec:rq3}

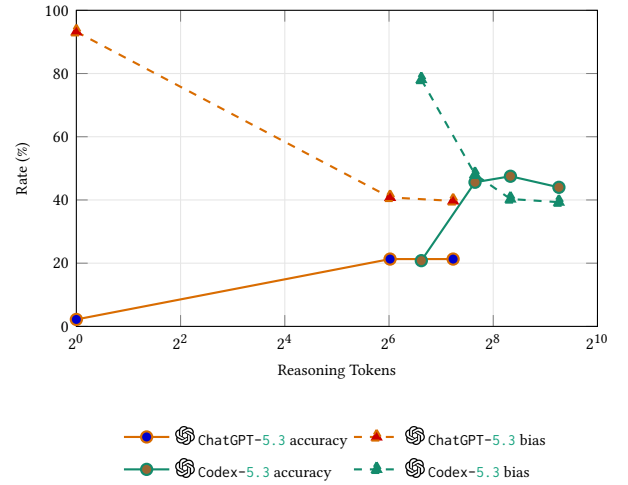
\begin{figure}[!ht]
\centering
\small
\begin{tikzpicture}
\begin{axis}[width=\columnwidth,height=0.68\columnwidth,xlabel={Reasoning Tokens},ylabel={Rate (\%)},xmin=0, xmax=10,ymin=0, ymax=100,xtick={0,2,4,6,8,10},xticklabels={$2^0$,$2^2$,$2^4$,$2^6$,$2^8$,$2^{10}$},ytick={0,20,40,60,80,100},grid=both,major grid style={draw=black!10},minor grid style={draw=black!5},tick label style={font=\scriptsize},label style={font=\scriptsize},legend style={font=\scriptsize, draw=none, at={(0.5,-0.28)}, anchor=north, legend columns=2},legend cell align=left,clip=false]
\addplot+[color=orange!85!black, thick, mark=*, mark size=2.1pt] coordinates {(0.00,2.2) (6.02,21.3) (7.23,21.3)};
\addlegendentry{\gptlogo~\gptchat accuracy}
\addplot+[color=orange!85!black, thick, dashed, mark=triangle*, mark size=2.7pt] coordinates {(0.00,93.2) (6.02,40.8) (7.23,39.7)};
\addlegendentry{\gptlogo~\gptchat bias}
\addplot+[color=gpt_green!85!black, thick, mark=*, mark size=2.1pt] coordinates {(6.62,20.8) (7.65,45.6) (8.33,47.5) (9.26,44.0)};
\addlegendentry{\gptlogo~\gptcodex accuracy}
\addplot+[color=gpt_green!85!black, thick, dashed, mark=triangle*, mark size=2.7pt] coordinates {(6.62,78.1) (7.65,48.1) (8.33,40.3) (9.26,39.3)};
\addlegendentry{\gptlogo~\gptcodex bias}
\end{axis}
\end{tikzpicture}
\caption{Reasoning effort \textit{vs} accuracy and bias rates for the two OpenAI models. Each point is a model-specific effort bin positioned by the $\log_2$ of its mean reasoning-token count. More reasoning consistently \rev{correlates with lower} bias.}
\label{fig:rq3-reasoning-effort}
\end{figure}

\begin{figure}[!ht]
\centering
\includegraphics[width=\columnwidth]{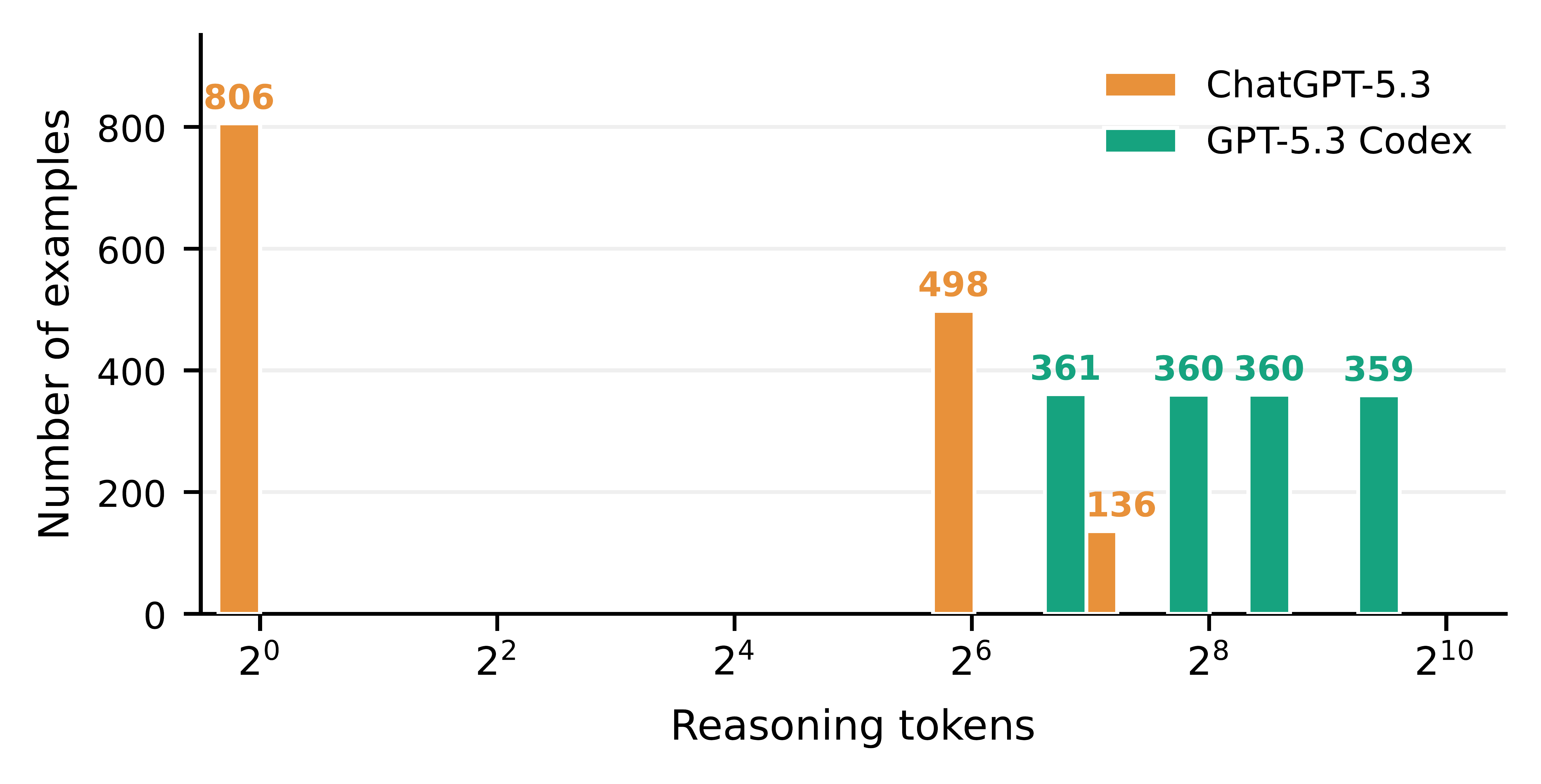}
\caption{Number of examples per reasoning effort: \gptlogo~\gptchat reasoning token distribution unbalanced: 806 of 1,440 examples use zero reasoning tokens. \gptlogo~\gptcodex always reasons, with bins of roughly equal size.}
\label{fig:rq3-bin-counts}
\end{figure}

Longer reasoning allows models to scrutinize the details, which may lead to better performance. For this analysis, we draw on the per-example token traces recorded in our inference logs. Only the two OpenAI models provide \texttt{usage\_reasoning\_tokens} values, so we restrict the RQ3 analysis to \gptlogo~\gptchat and \gptlogo~\gptcodex. \rev{We note that reasoning-token count is observed rather than experimentally controlled: models choose how much to reason per example. As such, the relationship we report is a correlation between reasoning effort and bias.}

Figures~\ref{fig:rq3-reasoning-effort} and~\ref{fig:rq3-bin-counts} tell a
clear story: \rev{\textit{examples answered with longer reasoning exhibit sharply lower bias, and accuracy improves up to a threshold where the gains diminish}}.

The contrast is sharpest for \gptlogo~\gptchat. When the model produces an
answer with zero reasoning tokens, which happens in over half of all examples (806 of 1,440; 56.0\%), it is almost always biased (93.2\%). Once reasoning kicks in at an intermediate level (centered at ${\sim}64$ tokens), accuracy rises to 21.3\% and bias is cut by more than half to 40.8\%. Beyond that threshold, however, performance plateaus. The highest-effort region (centered at ${\sim}149$ tokens) yields almost identical numbers (21.3\% accuracy, 39.7\% bias).

\gptlogo~\gptcodex shows a similar pattern, but with a more gradual increase. Accuracy rises from 20.8\% in the lowest-effort bin (centered at ${\sim}97$tokens) to 45.6--47.5\% across two intermediate bins (centered at ${\sim}200$
and ${\sim}321$ tokens), while bias retreats from 78.1\% to roughly 40\%. Yet the highest-effort bin (centered at ${\sim}611$ tokens) adds little: accuracy drops to 44.0\% and bias moves to 39.3\%. The takeaway is
consistent: \rev{examples with more deliberation land on the default bias answer far less often, but the extra effort does not guarantee better accuracy.}

% \anto{the type of token is a bit confusion, we should better clarify them}. 
% \khai{I'll remove this "type of token" part since it doesnt contribute to the narrative much}
% % Outcome-condition token counts add a useful nuance. Correct predictions do
% % consume more reasoning tokens than bias-aligned ones---70.7 vs.\ 21.3 for
% % \gptlogo~\gptchat and 337.0 vs.\ 259.4 for \gptlogo~\gptcodex---confirming
% % that getting it right takes genuine effort. The catch is that the
% % \emph{longest} reasoning traces tend to land in the ``Other'' error bucket
% % rather than in the correct one, particularly for \gptlogo~\gptcodex, where
% % Other-category responses average 446.3 reasoning tokens. Extended deliberation,
% % it seems, reliably pulls models away from reflexive pattern completion---but it
% % just as often redirects them toward a different wrong answer rather than toward
% % faithful visual grounding. This finding echoes~\citet{vlmsbiased}: thinking
% % longer can loosen the grip of bias, yet past a saturation point the extra
% % effort no longer converts into accuracy.

\begin{tcolorbox}[
    colback=white,
    colframe=black!35,
    boxrule=0.4pt,
    arc=4pt,
    left=8pt, right=8pt, top=8pt, bottom=8pt,
    title={\small Answer to RQ3},
    fonttitle=\bfseries,
    coltitle=white,
    colbacktitle=cyan!45!black,
]
\rev{Longer reasoning correlates with lower bias, but with diminishing returns for accuracy.} Intermediate effort drops bias from 93.2\% to $\sim$40\% for \gptlogo~\gptchat. General models like \gptlogo~\gptchat tend to \rev{jump directly} to the final answer, leading to stronger bias rate.
\end{tcolorbox}

\subsection{Qualitative analysis: examine model reasoning on modified patterns}
\label{sec:qualitative}

\begin{figure*}[!ht]
\centering
%--- Row 1: titles, images, ground truth ---
\begin{minipage}[t]{0.42\textwidth}
\textbf{(a) Self-talk-down: \claudelogo~\opus on a \texttt{80\%}-width card}\\[4pt]
\centering
\includegraphics[height=0.24\textheight,keepaspectratio]{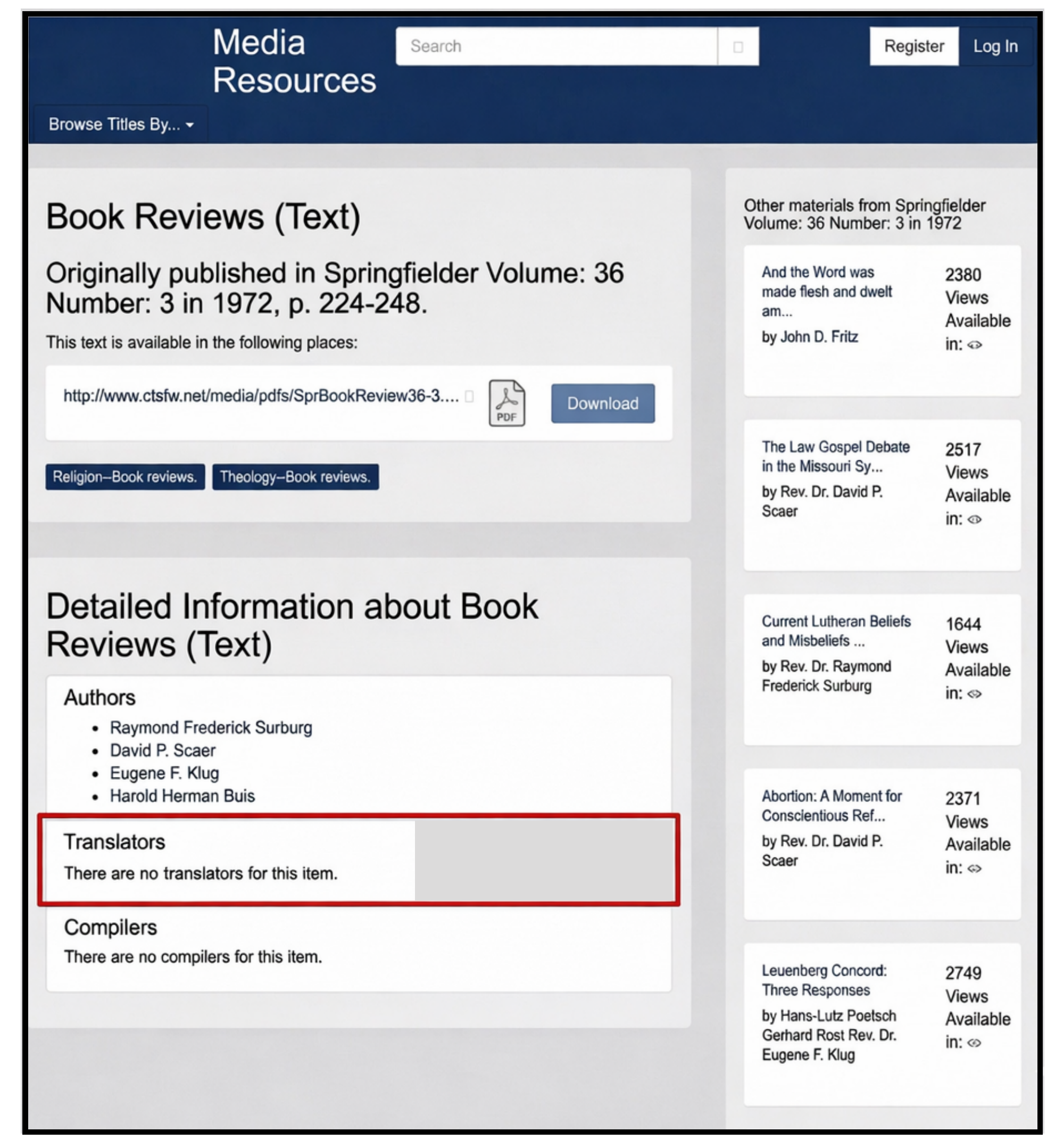}\\[2pt]
{Ground truth: \texttt{80\%}}
\end{minipage}%
\hfill
\begin{minipage}[t]{0.58\textwidth}
\textbf{(b) ``Despite the visual rendering'': \geminilogo~\geminiflash}\\[4pt]
\centering
\includegraphics[height=0.24\textheight,keepaspectratio]{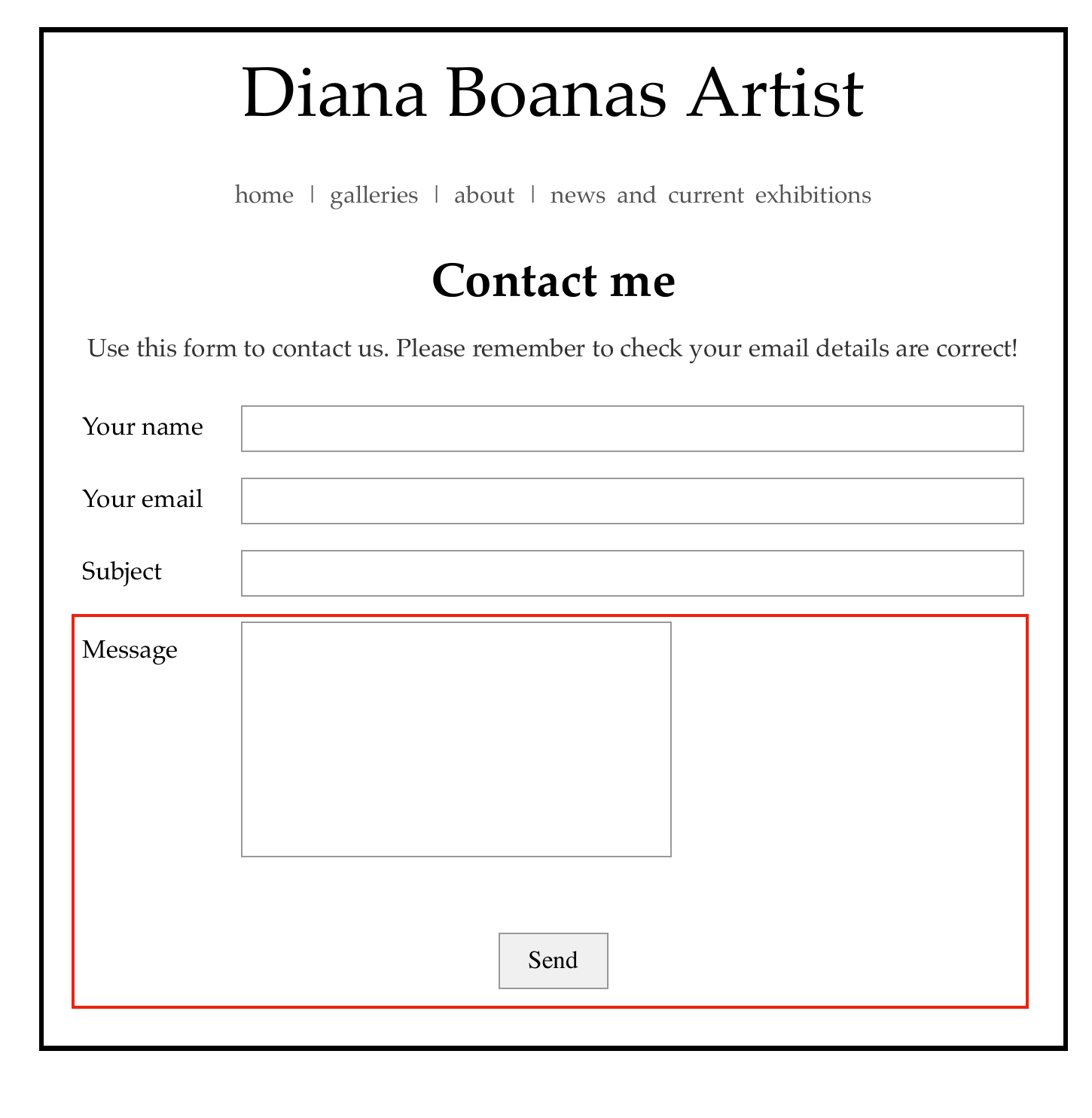}\\[2pt]
{Ground truth: \texttt{80\%}}
\end{minipage}

\vspace{6pt}

%--- Row 2: blue boxes (same fixed height, [t] aligned) ---
\noindent
\begin{minipage}[t]{0.42\textwidth}
\colorbox{blue!8}{\begin{minipage}[t][7em]{0.95\textwidth}
\textit{``The Translators card appears to be \textbf{about 80\% of the full width} compared to the other cards which span 100\%. [\ldots] However, examining more carefully [\ldots] the Translators card seems to extend to about the same width as the others. [\ldots] It's most likely also \textbf{100\% to match the consistent layout} of all four cards.''}
% \vfill
\hfill \underline{Answer}: \texttt{100\%}~\textcolor{red}{\textbf{$\boldsymbol{\times}$}}
\end{minipage}}
\end{minipage}%
\hfill
\begin{minipage}[t]{0.58\textwidth}
\colorbox{blue!8}{\begin{minipage}[t][7em]{0.95\textwidth}
\textit{``The `Message' field's input is \textbf{visibly narrower} than the input fields above it. [\ldots] it \textbf{spans a significantly smaller portion} of the row compared to the full-width fields above. [\ldots] \textbf{100\% is the most consistent choice despite the visual rendering} of the box inside.''}
\vfill
\hfill \underline{Answer}: \texttt{100\%}~\textcolor{red}{\textbf{$\boldsymbol{\times}$}}
\end{minipage}}
\end{minipage}
\caption{Two examples of pattern-completion bias in model reasoning. In (a), \claudelogo~\opus correctly estimates the card width at 80\% but overrides its own observation to match the repeated pattern. In (b), \geminilogo~\geminiflash explicitly acknowledges the visual deviation (``visibly narrower'', ``significantly smaller'') yet chooses the pattern-consistent value ``despite the visual rendering.'' Both models perceive the deviation but default to \texttt{100\%}.}
\label{fig:qualitative-examples}
\end{figure*}

To investigate why these models are systematically biased, we examine the free-text reasoning traces produced by the three models that expose them: \claudelogo~\sonnet, \claudelogo~\opus, and \geminilogo~\geminiflash. We omit the two OpenAI models as their reasoning tokens are encrypted. \rev{These traces come from the \emph{same} inference runs and the same prompt as the quantitative evaluation: the models emit their reasoning followed by the required JSON answer, and we score the JSON answer while analyzing the accompanying text.}

We identify a recurring three-step pattern that we call the \emph{pattern conformity} behavior: (1) the model observes the visual deviation, (2) it begins correct reasoning, and (3) it overrides its own observation to conform to the repeated pattern. We illustrate with representative examples from standard card screenshots.

\paragraph{Models perceive the deviation but override it.}
In many cases, models explicitly describe the visual anomaly and even estimate the correct value (Figures \ref{fig:teaser} \& \ref{fig:qualitative-examples}), then abandon their reasoning in favor of the pattern. For instance, on a \texttt{120\%}-width card perturbation (Figure \ref{fig:teaser}), \claudelogo~\sonnet computes the width ratio as $\sim$118\%, correctly identifies that rounding yields \texttt{120\%}, and then concludes: \textit{``However, looking more carefully, 100\% seems most consistent with the pattern''}. \claudelogo~\opus also describes a card as \textit{``about 80\% of the full width''} on a ground-truth \texttt{80\%} instance, then writes: \textit{``It's most likely also 100\% to match the consistent layout of all four cards.''} (Figure \ref{fig:qualitative-examples}a). In both cases, the model had the correct answer and discarded it. Similarly, \geminilogo~\geminiflash after noting that a card \textit{``spans a significantly smaller portion of the row''}, concludes \textit{``100\% is the most consistent choice despite the visual rendering of the box inside''} (Figure \ref{fig:qualitative-examples}b). These responses show that the bias operates as \rev{a} reasoning heuristic: models believe that repeated elements \emph{should} share the same value.

\subsection{\rev{Quantifying the Failure Modes}}
\label{sec:failure-modes}

\rev{While the qualitative examples in Section \ref{sec:qualitative} demonstrates how visual overrides happen in pattern-completion bias, they do not tell us how \emph{prevalent} each mechanism is. We therefore systematically classified every biased response that contains a free-text rationale, with 237 traces in total (192 from \claudelogo~\sonnet, 22 from \claudelogo~\opus, 23 from\geminilogo~\geminiflash), drawn from the same inference runs across both image conditions. We divide them into three mutually exclusive failure modes:}

\begin{itemize}
    \item \rev{\textbf{(A)~Code-anchored:} the rationale reasons only from the HTML source (\eg ``the other cards use \texttt{100\%}, so this should match'') and never references the rendered screenshot; the baseline answer is a default, not a visual judgment.}
    \item \rev{\textbf{(B)~Perceived-consistent:} the rationale cites the screenshot but reports the perturbed element as \emph{visually matching} its neighbors; the deviation is present in the pixels but is not registered.}
    \item \rev{\textbf{(C)~Observed-then-overrode:} the rationale explicitly notes that the element looks different (``wider,'' ``narrower,'' ``extends beyond''), yet the final answer is still the baseline. This behavior demonstrates strong pattern-completion bias in MLLMs which we illustrated in Figure~\ref{fig:qualitative-examples}.}
\end{itemize}

\rev{\paragraph{LLM-As-a-judge setup.} Labels are assigned by an LLM judge (\texttt{gpt-5.4-mini} via OpenRouter, temperature 0) using the two-step decision rule below. A key design choice is that the A/B/C discriminator is the trace's \emph{observation} of the pattern: phrases like ``should match the others'' appear in both B and C rationales, so category C requires an explicit claim that the element \emph{looks} different, even if the trace later talks itself back to ``consistent.''}

\begin{tcolorbox}[
    colback=white,
    colframe=black!35,
    boxrule=0.4pt,
    arc=4pt,
    left=8pt, right=8pt, top=6pt, bottom=6pt,
    title={\small \rev{Judge Prompt}},
    fonttitle=\bfseries,
    coltitle=white,
    colbacktitle=cyan!45!black,
]
\small \rev{Classify ONE reasoning trace into exactly one mode. Setup: the model filled a masked CSS width/font-size for one element among siblings; in the screenshot that element was genuinely perturbed to a DIFFERENT size, yet the model answered the baseline (100\%, same as siblings). Decide WHY, in strict order:

\smallskip
\noindent STEP 1 -- Does the trace describe the TARGET's RENDERED appearance (its size relative to siblings)? Talk purely about the HTML/CSS source (``the others are \texttt{width:100\%}, so match'') does NOT count.\\
\hspace*{1em}NO visual description $\rightarrow$ \textbf{\texttt{A (CODE\_ANCHORED)}}

\smallskip
\noindent STEP 2 -- (only if it looked) Does it assert that the target looks DIFFERENT -- narrower/wider/smaller/larger/bolder/overflows/``extends beyond'', or estimates a non-100 size (${\sim}80\%$, ${\sim}120\%$)?\\
\hspace*{1em}Only ever says SAME / consistent / uniform / matches $\rightarrow$\textbf{\texttt{ B (PERCEIVED\_CONSISTENT)}}\\
\hspace*{1em}Asserts a difference at any point, EVEN IF it then reconciles to ``same'' or picks 100\% to match the pattern $\rightarrow$ \textbf{\texttt{C (OBSERVED\_THEN\_OVERRODE)}}

\smallskip
\noindent CRITICAL: ``should match the others / be consistent'' is NOT evidence of C -- B says that too. C requires an explicit claim that the target LOOKS DIFFERENT.

\smallskip
\noindent Answer with ONE letter: A, B, or C. \\
\noindent ---\\
\noindent Model reasoning: \{\texttt{content}\}}
\end{tcolorbox}

\rev{\paragraph{Cross-validation.} We validated the judge against 30 traces sampled uniformly from the same population and hand-labeled \emph{before} inspecting the judge's output: agreement was 29/30 (96.7\%), and the single disagreement traced to a transcription error in the hand labels rather than a judge mistake.}

\begin{table}[!ht]
\centering
\caption{\rev{Failure-mode distribution over the 237 verbalized biased responses. Perceived-consistency (B) dominates; the explicit override (C) is real but smaller and model-dependent.}}
\label{tab:failure-modes}
\renewcommand{\arraystretch}{1.08}
\ifrevisions\color{blue}\fi
\resizebox{\columnwidth}{!}{%
\begin{tabular}{lcccc}
\toprule
Model & $n$ & A: code (\%) & B: perceived (\%) & C: override (\%) \\
\midrule
\claudelogo~\opus        & 22  & 13.6 & 40.9 & \textbf{45.5} \\
\claudelogo~\sonnet      & 192 & 26.0 & \textbf{61.5} & 12.5 \\
\geminilogo~\geminiflash & 23  & 39.1 & \textbf{56.5} & 4.3 \\
\midrule
\textbf{Overall}         & 237 & 26.2 & \textbf{59.1} & 14.8 \\
\bottomrule
\end{tabular}
}
\end{table}

\rev{Two conclusions follow from Table~\ref{tab:failure-modes}. First, the bias is not an artifact of ignoring the image: in $\sim$74\% of verbalized biased responses the model demonstrably engages the screenshot (B + C). Second, the dominant case is \emph{perceptual}: the pattern prior most often suppresses the perception of a deviation that is present in the pixels (B, 59.1\%), while the explicit override of Figure~\ref{fig:qualitative-examples} is a real but smaller, model-dependent mode (C, 14.8\% overall but 45.5\% for \claudelogo~\opus).}
\looseness=-1

\begin{tcolorbox}[
    colback=white,
    colframe=black!35,
    boxrule=0.4pt,
    arc=4pt,
    left=8pt, right=8pt, top=8pt, bottom=8pt,
    title={\small Takeaway from Failure Mode Analysis},
    fonttitle=\bfseries,
    coltitle=white,
    colbacktitle=cyan!45!black,
]
Analysis of the reasoning traces reveals that models (1) observe the visual deviation and begin correct inference, then (2) override their own observation to conform to the repeated pattern. \rev{Labeling of all 237 biased responses shows that $\sim$74\% engage the screenshot, with perceived-consistency (59.1\%) as the dominant mechanism and explicit override at 14.8\%.}
\looseness=-1
\end{tcolorbox}

\section{Threats to Validity}
\label{sec:threats}

\paragraph{Internal validity.}
Our benchmark relies on controlled synthetic construction: we inject a single localized deviation into otherwise pattern-consistent webpages and create noisy screenshot as synthetic transformations of the original renderings. This gives us precise control over visual context and makes it easier to attribute errors to pattern-completion bias. A similar design choice appears in prior multimodal benchmarks such as \textsc{HallusionBench}, \textsc{VALSE}, and \textsc{IllusionVQA}, which also use controlled perturbations to study specific visual failure modes. Likewise, screenshot-to-code benchmarks such as \textsc{Design2Code} and \textsc{WebSight} rely on curated or generated webpages to support consistency and reproducibility. At the same time, our synthetic setup does not fully capture the visual complexity of real-world web page screenshots, where unpredictable layout variations are admissible and may overstate or understate model susceptibility in practice. Extending the benchmark to more diverse web pages remains a key direction for future work.
\looseness=-1

\paragraph{Construct validity.}
Our primary metric is \textit{bias rate}, defined as the proportion of predictions
equal to \texttt{100\%}. This choice directly targets the failure mode under
study, pattern-completion bias, but it emphasizes one specific error
direction and does not fully capture other forms of visually grounded reasoning
failure. Furthermore, our task formulation constrains answers to percentage
values divisible by 10, which simplifies evaluation but may not reflect the
full range of CSS values a model would produce in unconstrained generation.
\looseness=-1

\paragraph{External validity.}
We evaluate two perturbation families---inline CSS width on card panels and
inline font-size on text sequences---across 30 \textsc{Design2Code} webpages.
Although this setting cleanly isolates pattern-completion bias, it does not
cover other localized deviations such as spacing, alignment, ordering, color,
or component presence/absence. All seed webpages originate from a single source
(\textsc{Design2Code}); benchmarks constructed from other web corpora, design
styles, or webpage structures may surface different bias magnitudes or patterns.
\rev{Moreover, our findings characterize \emph{one-shot, unaided} MLLM behavior:
in practice, developers and agentic systems may re-render generated code,
inspect computed styles, or apply visual-diff checks, and such tool-augmented
workflows may exhibit different bias profiles than the raw models we study.}

\paragraph{Conclusion validity.}
Our study reports results for five proprietary frontier MLLMs from three model
families. Although these represent the most widely adopted options in
professional development~\cite{stackoverflow2025technology_ai_models}, they do
not cover open-weight models or specialized screenshot-to-code systems.
Different architectures, training regimes, or alignment strategies may yield
different bias profiles. Accordingly, our findings should be read as evidence
that pattern-completion bias exists and follows a saliency-driven mechanism in
current frontier systems, not as an exhaustive characterization of all
multimodal code generation models.

\section{Discussion and Conclusion}
\label{sec:discussion}

We introduced a controlled benchmark for measuring \emph{visual
pattern-completion bias} in screenshot-to-code generation. The benchmark asks a
simple but important question: \textit{when a localized visual deviation conflicts with
a repeated UI pattern, does the model follow the pixels or the pattern?''} Across
720 base instances rendered under two matched image conditions (1,440
screenshots per model), the answer is unambiguous: all five MLLMs exhibit
pattern-completion bias, and its severity \rev{tracks} the strength of the visual signal (\ie saliency).
Card-width perturbations are detectable by the strongest model
(\gptlogo~\gptcodex at 68.61\% accuracy), but text font-size perturbations
cause even the best model to collapse (13.89\% accuracy, 70.56\% bias). Mean
text bias reaches 80.22\%, with \geminilogo~\geminiflash at 96.11\%. Noise,
subtler magnitudes, and boundary positions each further reduce saliency and
increase bias, confirming a single mechanism across every factor we tested.

These results impact and have implications along four axes:

\textbf{\textit{A saliency threshold \rev{shapes} when models can be trusted.}}
Pattern-completion bias operates along a gradient, not as a binary switch. When a localized deviation is spatially large, frontier models can sometimes
override their pattern prior. As the deviation shrinks below a perceptual
threshold, moving from layout-level card perturbations to character-level
font-size changes, models collapse to the dominant pattern at striking rates.
\textbf{For practitioners, this implies that screenshot-to-code output can be broadly trusted for coarse layout decisions but should \emph{not} be trusted for fine-grained styling without explicit verification}. The very properties that matter in polished front-end work (\eg subtle spacing, font sizing) sit below this threshold.

\textbf{\textit{End-to-end metrics can hide local grounding failures.}}
Standard screenshot-to-code benchmarks focus on overall visual similarity or structural overlap. A model can perform well on these metrics while still missing the small local deviations that our benchmark is designed to expose. In our setting, this failure is not random: models often revert to the repeated baseline \texttt{100\%}. Because the resulting outputs still look globally plausible, these errors may not be captured by page-level metrics. This suggests the need for complementary evaluation settings that use controlled edits and localized perturbations, so grounding failures can be observed directly.

\textbf{\textit{Toward mitigation: prompting, verification, and training.}}
Our findings point to several complementary directions for reducing
pattern-completion bias in practice. First, the RQ3 results show that even
moderate reasoning effort halves bias for \gptlogo~\model{ChatGPT -\textcolor{gpt_green}{5.3}} (93.2\%
$\rightarrow$ 40.8\%), \textbf{suggesting that prompts which explicitly instruct the
model to examine each element individually, rather than requesting a single
holistic completion, could reduce default-to-pattern behavior}. Our qualitative
analysis (Section~\ref{sec:qualitative}) shows that models already perform this
reasoning internally but then override it; prompts that penalize or flag
self-contradiction in the reasoning trace could help close this gap. Second, a
lightweight post-generation verification pass, re-rendering the generated code
and comparing the target element's computed CSS value against the
screenshot, could catch the most egregious errors without requiring changes to
the model itself. Third, longer-term training interventions such as
reinforcement learning from visual feedback, where the reward signal is derived
from pixel-level comparison of the generated code's rendering against the input
screenshot, could directly penalize outputs that are pattern consistent but visually incorrect. Curriculum learning \cite{bengio2009curriculum} as a training technique that exposes models to progressively subtler deviations during fine-tuning may also help lower the saliency threshold at which grounding breaks down.
\looseness=-1

\textbf{\textit{Implications for professional adoption.}}
Screenshot-to-code tools are no longer experimental---they are embedded in the
workflows of a large fraction of professional developers. Our findings imply
that developers using these tools for generating code from pictures depicting real web components,  should treat
the generated output as a \emph{draft} that preserves global structure but may
silently normalize fine-grained styling decisions. This, holds relevance, especially in contexts where pixel-level fidelity matters, such as design
systems with strict spacing tokens, accessibility-sensitive layouts, or
brand-compliant interfaces. Teams relying on screenshot-to-code pipelines
would benefit from integrating automated visual regression checks that flag
discrepancies between the input screenshot and the rendered output at the
element level.

\medskip
\noindent Looking ahead, the pattern conformity phenomenon, where models derive
the correct answer and then discard it in favor of pattern conformity, points
to a deeper tension in how current \mllms{} balance prior knowledge against input
evidence. Our benchmark provides the first controlled, reproducible setting in which
this tension can be measured and tracked as models improve. 
As screenshot-to-code technology matures and software engineering workflows grow increasingly automated, we hope this framework encourages testing procedures that push multimodal models toward faithful visual understanding, capable of operating on entire codebases, rather than toward more confident pattern completion.

\begin{acks}
This work was supported in part by NSF grant IIS-2533367 and NSF grant CCF-245105. The views expressed herein are the authors' own and do not necessarily reflect those of the sponsors.
\end{acks}

\section*{Data Availability}
We made all study artifacts, including the dataset, source code, and documentation, publicly accessible on Zenodo ~\cite{rep_package}.

%\pagebreak
%\newpage
\balance
\bibliographystyle{ACM-Reference-Format}
\bibliography{util/main}

@article{jiang2025viscodex,
  title={Viscodex: Unified multimodal code generation via merging vision and coding models},
  author={Jiang, Lingjie and Huang, Shaohan and Wu, Xun and Li, Yixia and Zhang, Dongdong and Wei, Furu},
  journal={arXiv preprint arXiv:2508.09945},
  year={2025}
}

@article{jiang2025screencoder,
  title={Screencoder: Advancing visual-to-code generation for front-end automation via modular multimodal agents},
  author={Jiang, Yilei and Zheng, Yaozhi and Wan, Yuxuan and Han, Jiaming and Wang, Qunzhong and Lyu, Michael R and Yue, Xiangyu},
  journal={arXiv preprint arXiv:2507.22827},
  year={2025}
}

@article{wu2024uicoder,
  title        = {UICoder: Finetuning Large Language Models to Generate User Interface Code through Automated Feedback},
  author       = {Wu, Jason and Schoop, Eldon and Leung, Alan and Barik, Titus and Bigham, Jeffrey P. and Nichols, Jeffrey},
  journal      = {arXiv preprint arXiv:2406.07739},
  year         = {2024},
  url          = {https://arxiv.org/abs/2406.07739}
}

@article{xiao2024prototype2code,
  title        = {Prototype2Code: End-to-end Front-end Code Generation from UI Design Prototypes},
  author       = {Xiao, Shuhong and Chen, Yunnong and Li, Jiazhi and Chen, Liuqing and Sun, Lingyun and Zhou, Tingting},
  journal      = {arXiv preprint arXiv:2405.04975},
  year         = {2024},
  url          = {https://arxiv.org/abs/2405.04975}
}

@article{gui2025uicopilot,
  title        = {UICopilot: Automating UI Synthesis via Hierarchical Code Generation from Webpage Designs},
  author       = {Gui, Yi and Wan, Yao and Li, Zhen and Zhang, Zhongyi and Chen, Dongping and Zhang, Hongyu and Su, Yi and Chen, Bohua and Zhou, Xing and Jiang, Wenbin and Zhang, Xiangliang},
  journal      = {arXiv preprint arXiv:2505.09904},
  year         = {2025},
  url          = {https://arxiv.org/abs/2505.09904}
}

@article{yang2025ui2code,
  title={UI2Code\^{} N: A Visual Language Model for Test-Time Scalable Interactive UI-to-Code Generation},
  author={Yang, Zhen and Hong, Wenyi and Xu, Mingde and Fan, Xinyue and Wang, Weihan and Cheng, Jiele and Gu, Xiaotao and Tang, Jie},
  journal={arXiv preprint arXiv:2511.08195},
  year={2025}
}

@inproceedings{lee2023pix2struct,
  title={Pix2struct: Screenshot parsing as pretraining for visual language understanding},
  author={Lee, Kenton and Joshi, Mandar and Turc, Iulia Raluca and Hu, Hexiang and Liu, Fangyu and Eisenschlos, Julian Martin and Khandelwal, Urvashi and Shaw, Peter and Chang, Ming-Wei and Toutanova, Kristina},
  booktitle={International Conference on Machine Learning},
  pages={18893--18912},
  year={2023},
  organization={PMLR}
}

@article{zhou2024declarui,
  title        = {Bridging Design and Development with Automated Declarative UI Code Generation},
  author       = {Zhou, Ting and Zhao, Yanjie and Hou, Xinyi and Sun, Xiaoyu and Chen, Kai and Wang, Haoyu},
  journal      = {arXiv preprint arXiv:2409.11667},
  year         = {2024},
  url          = {https://arxiv.org/abs/2409.11667}
}

@article{wan2024dcgen,
  title        = {Automatically Generating UI Code from Screenshot: A Divide-and-Conquer-Based Approach},
  author       = {Wan, Yuxuan and Wang, Chaozheng and Dong, Yi and Wang, Wenxuan and Li, Shuqing and Huo, Yintong and Lyu, Michael R.},
  journal      = {arXiv preprint arXiv:2406.16386},
  year         = {2024},
  url          = {https://arxiv.org/abs/2406.16386}
}

@inproceedings{si2025design2code,
  title={Design2code: Benchmarking multimodal code generation for automated front-end engineering},
  author={Si, Chenglei and Zhang, Yanzhe and Li, Ryan and Yang, Zhengyuan and Liu, Ruibo and Yang, Diyi},
  booktitle={Proceedings of the 2025 Conference of the Nations of the Americas Chapter of the Association for Computational Linguistics: Human Language Technologies (Volume 1: Long Papers)},
  pages={3956--3974},
  year={2025}
}

@misc{rep_package,
  title        = {Replication Package.},
  url          = {https://doi.org/10.5281/zenodo.19341952},
}

@inproceedings{awal2025webmmu,
  title={Webmmu: A benchmark for multimodal multilingual website understanding and code generation},
  author={Awal, Rabiul and Massoud, Mahsa and Feizi, Aarash and Li, Zichao and Wang, Suyuchen and Pal, Christopher and Agrawal, Aishwarya and Vazquez, David and Reddy, Siva and Rodriguez, Juan A and others},
  booktitle={Proceedings of the 2025 Conference on Empirical Methods in Natural Language Processing},
  pages={25129--25156},
  year={2025}
}

@inproceedings{bengio2009curriculum,
  title={Curriculum learning},
  author={Bengio, Yoshua and Louradour, J{\'e}r{\^o}me and Collobert, Ronan and Weston, Jason},
  booktitle={Proceedings of the 26th annual international conference on machine learning},
  pages={41--48},
  year={2009}
}

@article{yun2024web2code,
  title={Web2code: A large-scale webpage-to-code dataset and evaluation framework for multimodal llms},
  author={Yun, Sukmin and Thushara, Rusiru and Bhat, Mohammad and Wang, Yongxin and Deng, Mingkai and Wang, Jinhong and Tao, Tianhua and Li, Junbo and Li, Haonan and Nakov, Preslav and others},
  journal={Advances in neural information processing systems},
  volume={37},
  pages={112134--112157},
  year={2024}
}

@inproceedings{beltramelli2018pix2code,
  title={pix2code: Generating code from a graphical user interface screenshot},
  author={Beltramelli, Tony},
  booktitle={Proceedings of the ACM SIGCHI symposium on engineering interactive computing systems},
  pages={1--6},
  year={2018}
}

@article{gui2024webcode2m,
  title={WebCode2M: A Real-World Dataset for Code Generation from Webpage Designs},
  author={Gui, Yi and Li, Zhen and Wan, Yao and Shi, Yemin and Zhang, Hongyu and Su, Yi and Chen, Bohua and Wu, Siyuan and Zhou, Xing and Jiang, Wenbin and Jin, Hai and Zhang, Xiangliang},
  journal={arXiv preprint arXiv:2404.06369},
  year={2024},
  url={https://arxiv.org/abs/2404.06369}
}

@article{lin2025webuibench,
  title={WebUIBench: A Comprehensive Benchmark for Evaluating Multimodal Large Language Models in WebUI-to-Code},
  author={Lin, Zhiyu and Zhou, Zhengda and Zhao, Zhiyuan and Wan, Tianrui and Ma, Yilun and Gao, Junyu and Li, Xuelong},
  journal={arXiv preprint arXiv:2506.07818},
  year={2025},
  url={https://arxiv.org/abs/2506.07818}
}

@article{xiao2025designbench,
  title={DesignBench: A Comprehensive Benchmark for MLLM-based Front-end Code Generation},
  author={Xiao, Jingyu and Wang, Ming and Lam, Man Ho and Wan, Yuxuan and Liu, Junliang and Huo, Yintong and Lyu, Michael R.},
  journal={arXiv preprint arXiv:2506.06251},
  year={2025},
  url={https://arxiv.org/abs/2506.06251}
}

@article{zhu2025frontendbench,
  title={FrontendBench: A Benchmark for Evaluating LLMs on Front-End Development via Automatic Evaluation},
  author={Zhu, Hongda and Zhang, Yiwen and Zhao, Bing and Ding, Jingzhe and Liu, Siyao and Liu, Tong and Wang, Dandan and Liu, Yanan and Li, Zhaojian},
  journal={arXiv preprint arXiv:2506.13832},
  year={2025},
  url={https://arxiv.org/abs/2506.13832}
}

@article{sun2025fullfront,
  title={FullFront: Benchmarking MLLMs Across the Full Front-End Engineering Workflow},
  author={Sun, Haoyu and Wang, Huichen Will and Gu, Jiawei and Li, Linjie and Cheng, Yu},
  journal={arXiv preprint arXiv:2505.17399},
  year={2025},
  url={https://arxiv.org/abs/2505.17399}
}

@article{xiao2024interaction2code,
  title={Interaction2Code: Benchmarking MLLM-based Interactive Webpage Code Generation from Interactive Prototyping},
  author={Xiao, Jingyu and Wan, Yuxuan and Huo, Yintong and Wang, Zixin and Xu, Xinyi and Wang, Wenxuan and Xu, Zhiyao and Wang, Yuhang and Lyu, Michael R.},
  journal={arXiv preprint arXiv:2411.03292},
  year={2024},
  url={https://arxiv.org/abs/2411.03292}
}

@article{laurencon2024websight,
  title={Unlocking the Conversion of Web Screenshots into HTML Code with the WebSight Dataset},
  author={Lauren{\c{c}}on, Hugo and Tronchon, L{\'e}o and Sanh, Victor},
  journal={arXiv preprint arXiv:2403.09029},
  year={2024},
  url={https://arxiv.org/abs/2403.09029}
}

@article{vlmsbiased,
  author       = {An Vo and
                  Khai{-}Nguyen Nguyen and
                  Mohammad Reza Taesiri and
                  Vy Tuong Dang and
                  Anh Totti Nguyen and
                  Daeyoung Kim},
  title        = {Vision Language Models are Biased},
  journal      = {CoRR},
  volume       = {abs/2505.23941},
  year         = {2025},
  url          = {https://doi.org/10.48550/arXiv.2505.23941},
  doi          = {10.48550/ARXIV.2505.23941},
  eprinttype    = {arXiv},
  eprint       = {2505.23941},
  bibsource    = {dblp computer science bibliography, https://dblp.org}
}

@inproceedings{huang2024visual,
  author       = {Wen Huang and
                  Hongbin Liu and
                  Minxin Guo and
                  Neil Gong},
  editor       = {Lun{-}Wei Ku and
                  Andre Martins and
                  Vivek Srikumar},
  title        = {Visual Hallucinations of Multi-modal Large Language Models},
  booktitle    = {Findings of the Association for Computational Linguistics, {ACL} 2024,
                  Bangkok, Thailand and virtual meeting, August 11-16, 2024},
  pages        = {9614--9631},
  publisher    = {Association for Computational Linguistics},
  year         = {2024},
  url          = {https://doi.org/10.18653/v1/2024.findings-acl.573},
  doi          = {10.18653/V1/2024.FINDINGS-ACL.573},
  bibsource    = {dblp computer science bibliography, https://dblp.org}
}

@inproceedings{tong2024eyes,
  title={Eyes wide shut? exploring the visual shortcomings of multimodal llms},
  author={Tong, Shengbang and Liu, Zhuang and Zhai, Yuexiang and Ma, Yi and LeCun, Yann and Xie, Saining},
  booktitle={CVPR},
  year={2024}
}

@inproceedings{ye2024beaf,
  title={Beaf: Observing before-after changes to evaluate hallucination in vision-language models},
  author={Ye-Bin, Moon and Hyeon-Woo, Nam and Choi, Wonseok and Oh, Tae-Hyun},
  booktitle={European Conference on Computer Vision},
  pages={232--248},
  year={2024},
  organization={Springer}
}

@inproceedings{parcalabescu-etal-2022-valse,
    title = "{VALSE}: A Task-Independent Benchmark for Vision and Language Models Centered on Linguistic Phenomena",
    author = "Parcalabescu, Letitia  and
      Cafagna, Michele  and
      Muradjan, Lilitta  and
      Frank, Anette  and
      Calixto, Iacer  and
      Gatt, Albert",
    editor = "Muresan, Smaranda  and
      Nakov, Preslav  and
      Villavicencio, Aline",
    booktitle = "Proceedings of the 60th Annual Meeting of the Association for Computational Linguistics (Volume 1: Long Papers)",
    month = may,
    year = "2022",
    address = "Dublin, Ireland",
    publisher = "Association for Computational Linguistics",
    url = "https://aclanthology.org/2022.acl-long.567/",
    doi = "10.18653/v1/2022.acl-long.567",
    pages = "8253--8280",
}

@article{shahgir2024illusionvqa,
  title={IllusionVQA: A challenging optical illusion dataset for vision language models},
  author={Shahgir, Haz Sameen and Sayeed, Khondker Salman and Bhattacharjee, Abhik and Ahmad, Wasi Uddin and Dong, Yue and Shahriyar, Rifat},
  journal={arXiv preprint arXiv:2403.15952},
  year={2024}
}

@inproceedings{bitton2023breaking,
  title={Breaking common sense: Whoops! a vision-and-language benchmark of synthetic and compositional images},
  author={Bitton-Guetta, Nitzan and Bitton, Yonatan and Hessel, Jack and Schmidt, Ludwig and Elovici, Yuval and Stanovsky, Gabriel and Schwartz, Roy},
  booktitle={Proceedings of the IEEE/CVF International Conference on Computer Vision},
  pages={2616--2627},
  year={2023}
}

@inproceedings{
rome,
title={{ROME}: Evaluating Pre-trained Vision-Language Models on Reasoning beyond Visual Common Sense},
author={Kankan Zhou and Eason Lai and Wei Bin Au Yeong and Kyriakos Mouratidis and Jing Jiang},
booktitle={Findings of the Association for Computational Linguistics: EMNLP},
year={2023},
url={https://openreview.net/forum?id=N6sXsHuWDE}
}

@inproceedings{guan2024hallusionbench,
  title={Hallusionbench: an advanced diagnostic suite for entangled language hallucination and visual illusion in large vision-language models},
  author={Guan, Tianrui and Liu, Fuxiao and Wu, Xiyang and Xian, Ruiqi and Li, Zongxia and Liu, Xiaoyu and Wang, Xijun and Chen, Lichang and Huang, Furong and Yacoob, Yaser and others},
  booktitle={Proceedings of the IEEE/CVF Conference on Computer Vision and Pattern Recognition},
  pages={14375--14385},
  year={2024}
}

@article{liu2024phd,
  title={Phd: A chatgpt-prompted visual hallucination evaluation dataset},
  author={Liu, Jiazhen and Fu, Yuhan and Xie, Ruobing and Xie, Runquan and Sun, Xingwu and Lian, Fengzong and Kang, Zhanhui and Li, Xirong},
  journal={arXiv preprint arXiv:2403.11116},
  year={2024}
}

@inproceedings{lee2025vlind,
  title={Vlind-bench: Measuring language priors in large vision-language models},
  author={Lee, Kang-il and Kim, Minbeom and Yoon, Seunghyun and Kim, Minsung and Lee, Dongryeol and Koh, Hyukhun and Jung, Kyomin},
  booktitle={Findings of the Association for Computational Linguistics: NAACL 2025},
  pages={4129--4144},
  year={2025}
}

@online{codex_overview,
  author       = {{OpenAI}},
  title        = {https://openai.com/index/introducing-codex/},
  year         = {2025},
  url          = {https://openai.com/index/introducing-codex/},
}

@online{anthropic_claude_code_overview,
  author       = {{Anthropic}},
  title        = {Claude Code overview},
  year         = {2026},
  url          = {https://code.claude.com/docs/en/overview},
  note         = {Claude Code Docs. Accessed: 2026-03-09}
}

@online{google_gemini_code_assist_overview,
  author       = {{Google}},
  title        = {Gemini Code Assist overview},
  year         = {2026},
  url          = {https://developers.google.com/gemini-code-assist/docs/overview},
  note         = {Google Developers Documentation. Accessed: 2026-03-09}
}

@online{stackoverflow2025technology_ai_models,
  author       = {{Stack Overflow}},
  title        = {Technology --- 2025 Stack Overflow Developer Survey},
  year         = {2025},
  url          = {https://survey.stackoverflow.co/2025/technology#most-popular-technologies-ai-models-ai-models-prof},
  note         = {Accessed: 2026-03-20}
}

@online{openrouter,
  author       = {{OpenRouter}},
  title        = {OpenRouter},
  year         = {2026},
  url          = {https://openrouter.ai/},
  note         = {Accessed: 2026-03-20}
}

@misc{openai2026gpt53codex,
  author       = {{OpenAI}},
  title        = {Introducing {GPT-5.3-Codex}},
  year         = {2026},
  month        = feb,
  howpublished = {\url{https://openai.com/index/introducing-gpt-5-3-codex/}},
  url          = {https://openai.com/index/introducing-gpt-5-3-codex/},
  note         = {Accessed: 2026-03-26}
}

@misc{openai2026gpt53instant,
  author       = {{OpenAI}},
  title        = {{GPT-5.3} Instant: Smoother, More Useful Everyday Conversations},
  year         = {2026},
  month        = mar,
  howpublished = {\url{https://openai.com/index/gpt-5-3-instant/}},
  url          = {https://openai.com/index/gpt-5-3-instant/},
  note         = {Accessed: 2026-03-26}
}

@misc{anthropic2026opus46,
  author       = {{Anthropic}},
  title        = {Introducing {Claude Opus 4.6}},
  year         = {2026},
  month        = feb,
  howpublished = {\url{https://www.anthropic.com/news/claude-opus-4-6}},
  url          = {https://www.anthropic.com/news/claude-opus-4-6},
  note         = {Accessed: 2026-03-26}
}

@misc{anthropic2026sonnet46,
  author       = {{Anthropic}},
  title        = {Introducing {Claude Sonnet 4.6}},
  year         = {2026},
  month        = feb,
  howpublished = {\url{https://www.anthropic.com/news/claude-sonnet-4-6}},
  url          = {https://www.anthropic.com/news/claude-sonnet-4-6},
  note         = {Accessed: 2026-03-26}
}

@misc{google2025gemini3flash,
  author       = {{Google DeepMind}},
  title        = {{Gemini 3 Flash}: Frontier Intelligence Built for Speed},
  year         = {2025},
  month        = dec,
  howpublished = {\url{https://blog.google/products/gemini/gemini-3-flash/}},
  url          = {https://blog.google/products/gemini/gemini-3-flash/},
  note         = {Accessed: 2026-03-26}
}

@article{chen2021evaluating,
  title={Evaluating large language models trained on code},
  author={Chen, Mark and Tworek, Jerry and Jun, Heewoo and Yuan, Qiming and Pinto, Henrique Ponde De Oliveira and Kaplan, Jared and Edwards, Harri and Burda, Yuri and Joseph, Nicholas and Brockman, Greg and others},
  journal={arXiv preprint arXiv:2107.03374},
  year={2021}
}

@inproceedings{xia2023automated,
  title={Automated program repair in the era of large pre-trained language models},
  author={Xia, Chunqiu Steven and Wei, Yuxiang and Zhang, Lingming},
  booktitle={2023 IEEE/ACM 45th International Conference on Software Engineering (ICSE)},
  pages={1482--1494},
  year={2023},
  organization={IEEE}
}

@article{hou2024large,
  title={Large language models for software engineering: A systematic literature review},
  author={Hou, Xinyi and Zhao, Yanjie and Liu, Yue and Yang, Zhou and Wang, Kailong and Li, Li and Luo, Xiapu and Lo, David and Grundy, John and Wang, Haoyu},
  journal={ACM Transactions on Software Engineering and Methodology},
  volume={33},
  number={8},
  pages={1--79},
  year={2024},
  publisher={ACM New York, NY}
}

@incollection{white2024chatgpt,
  title={Chatgpt prompt patterns for improving code quality, refactoring, requirements elicitation, and software design},
  author={White, Jules and Hays, Sam and Fu, Quchen and Spencer-Smith, Jesse and Schmidt, Douglas C},
  booktitle={Generative ai for effective software development},
  pages={71--108},
  year={2024},
  publisher={Springer}
}

@article{wang2024software,
  title={Software testing with large language models: Survey, landscape, and vision},
  author={Wang, Junjie and Huang, Yuchao and Chen, Chunyang and Liu, Zhe and Wang, Song and Wang, Qing},
  journal={IEEE Transactions on Software Engineering},
  volume={50},
  number={4},
  pages={911--936},
  year={2024},
  publisher={IEEE}
}

@misc{huang2025biastestingmitigationllmbased,
      title={Bias Testing and Mitigation in LLM-based Code Generation}, 
      author={Dong Huang and Jie M. Zhang and Qingwen Bu and Xiaofei Xie and Junjie Chen and Heming Cui},
      year={2025},
      eprint={2309.14345},
      archivePrefix={arXiv},
      primaryClass={cs.SE},
      url={https://arxiv.org/abs/2309.14345}, 
}

@article{liu2023uncovering,
  title={Uncovering and quantifying social biases in code generation},
  author={Liu, Yan and Chen, Xiaokang and Gao, Yan and Su, Zhe and Zhang, Fengji and Zan, Daoguang and Lou, Jian-Guang and Chen, Pin-Yu and Ho, Tsung-Yi},
  journal={Advances in Neural Information Processing Systems},
  volume={36},
  pages={2368--2380},
  year={2023}
}

@misc{ling2025biasunveiledinvestigatingsocial,
      title={Bias Unveiled: Investigating Social Bias in LLM-Generated Code}, 
      author={Lin Ling and Fazle Rabbi and Song Wang and Jinqiu Yang},
      year={2025},
      eprint={2411.10351},
      archivePrefix={arXiv},
      primaryClass={cs.SE},
      url={https://arxiv.org/abs/2411.10351}, 
}

@misc{playwright2020,
  author       = {Microsoft},
  title        = {Playwright},
  year         = {2020},
  howpublished = {\url{https://github.com/microsoft/playwright}}
}
\end{document}